\documentclass[runningheads]{llncs}
\usepackage{graphicx}
\usepackage{color}
\usepackage{amsmath}
\usepackage{amssymb}
\usepackage{booktabs}
\usepackage{xcolor}
\usepackage{tikz}
\usepackage{quantikz}
\usepackage[draft]{minted} 
\usepackage[square,numbers]{natbib}
\usepackage{subcaption}
\usepackage{listings}
\setminted{linenos=true}
\definecolor{darkgreen}{RGB}{0,128,0}
\definecolor{darkred}{RGB}{128,0,0}
\definecolor{darkblue}{RGB}{0,0,128}

\newcommand{\Na}[0]{\mathbb{N}}
\newcommand{\Za}[0]{\mathbb{Z}}

\newcommand{\Bo}[0]{\{0,1\}}

\makeatletter
\def\PYG@reset{\let\PYG@it=\relax \let\PYG@bf=\relax%
    \let\PYG@ul=\relax \let\PYG@tc=\relax%
    \let\PYG@bc=\relax \let\PYG@ff=\relax}
\def\PYG@tok#1{\csname PYG@tok@#1\endcsname}
\def\PYG@toks#1+{\ifx\relax#1\empty\else%
    \PYG@tok{#1}\expandafter\PYG@toks\fi}
\def\PYG@do#1{\PYG@bc{\PYG@tc{\PYG@ul{%
    \PYG@it{\PYG@bf{\PYG@ff{#1}}}}}}}
\def\PYG#1#2{\PYG@reset\PYG@toks#1+\relax+\PYG@do{#2}}

\expandafter\def\csname PYG@tok@w\endcsname{\def\PYG@tc##1{\textcolor[rgb]{0.73,0.73,0.73}{##1}}}
\expandafter\def\csname PYG@tok@c\endcsname{\let\PYG@it=\textit\def\PYG@tc##1{\textcolor[rgb]{0.25,0.50,0.50}{##1}}}
\expandafter\def\csname PYG@tok@cp\endcsname{\def\PYG@tc##1{\textcolor[rgb]{0.74,0.48,0.00}{##1}}}
\expandafter\def\csname PYG@tok@k\endcsname{\let\PYG@bf=\textbf\def\PYG@tc##1{\textcolor[rgb]{0.00,0.50,0.00}{##1}}}
\expandafter\def\csname PYG@tok@kp\endcsname{\def\PYG@tc##1{\textcolor[rgb]{0.00,0.50,0.00}{##1}}}
\expandafter\def\csname PYG@tok@kt\endcsname{\def\PYG@tc##1{\textcolor[rgb]{0.69,0.00,0.25}{##1}}}
\expandafter\def\csname PYG@tok@o\endcsname{\def\PYG@tc##1{\textcolor[rgb]{0.40,0.40,0.40}{##1}}}
\expandafter\def\csname PYG@tok@ow\endcsname{\let\PYG@bf=\textbf\def\PYG@tc##1{\textcolor[rgb]{0.67,0.13,1.00}{##1}}}
\expandafter\def\csname PYG@tok@nb\endcsname{\def\PYG@tc##1{\textcolor[rgb]{0.00,0.50,0.00}{##1}}}
\expandafter\def\csname PYG@tok@nf\endcsname{\def\PYG@tc##1{\textcolor[rgb]{0.00,0.00,1.00}{##1}}}
\expandafter\def\csname PYG@tok@nc\endcsname{\let\PYG@bf=\textbf\def\PYG@tc##1{\textcolor[rgb]{0.00,0.00,1.00}{##1}}}
\expandafter\def\csname PYG@tok@nn\endcsname{\let\PYG@bf=\textbf\def\PYG@tc##1{\textcolor[rgb]{0.00,0.00,1.00}{##1}}}
\expandafter\def\csname PYG@tok@ne\endcsname{\let\PYG@bf=\textbf\def\PYG@tc##1{\textcolor[rgb]{0.82,0.25,0.23}{##1}}}
\expandafter\def\csname PYG@tok@nv\endcsname{\def\PYG@tc##1{\textcolor[rgb]{0.10,0.09,0.49}{##1}}}
\expandafter\def\csname PYG@tok@no\endcsname{\def\PYG@tc##1{\textcolor[rgb]{0.53,0.00,0.00}{##1}}}
\expandafter\def\csname PYG@tok@nl\endcsname{\def\PYG@tc##1{\textcolor[rgb]{0.63,0.63,0.00}{##1}}}
\expandafter\def\csname PYG@tok@ni\endcsname{\let\PYG@bf=\textbf\def\PYG@tc##1{\textcolor[rgb]{0.60,0.60,0.60}{##1}}}
\expandafter\def\csname PYG@tok@na\endcsname{\def\PYG@tc##1{\textcolor[rgb]{0.49,0.56,0.16}{##1}}}
\expandafter\def\csname PYG@tok@nt\endcsname{\let\PYG@bf=\textbf\def\PYG@tc##1{\textcolor[rgb]{0.00,0.50,0.00}{##1}}}
\expandafter\def\csname PYG@tok@nd\endcsname{\def\PYG@tc##1{\textcolor[rgb]{0.67,0.13,1.00}{##1}}}
\expandafter\def\csname PYG@tok@s\endcsname{\def\PYG@tc##1{\textcolor[rgb]{0.73,0.13,0.13}{##1}}}
\expandafter\def\csname PYG@tok@sd\endcsname{\let\PYG@it=\textit\def\PYG@tc##1{\textcolor[rgb]{0.73,0.13,0.13}{##1}}}
\expandafter\def\csname PYG@tok@si\endcsname{\let\PYG@bf=\textbf\def\PYG@tc##1{\textcolor[rgb]{0.73,0.40,0.53}{##1}}}
\expandafter\def\csname PYG@tok@se\endcsname{\let\PYG@bf=\textbf\def\PYG@tc##1{\textcolor[rgb]{0.73,0.40,0.13}{##1}}}
\expandafter\def\csname PYG@tok@sr\endcsname{\def\PYG@tc##1{\textcolor[rgb]{0.73,0.40,0.53}{##1}}}
\expandafter\def\csname PYG@tok@ss\endcsname{\def\PYG@tc##1{\textcolor[rgb]{0.10,0.09,0.49}{##1}}}
\expandafter\def\csname PYG@tok@sx\endcsname{\def\PYG@tc##1{\textcolor[rgb]{0.00,0.50,0.00}{##1}}}
\expandafter\def\csname PYG@tok@m\endcsname{\def\PYG@tc##1{\textcolor[rgb]{0.40,0.40,0.40}{##1}}}
\expandafter\def\csname PYG@tok@gh\endcsname{\let\PYG@bf=\textbf\def\PYG@tc##1{\textcolor[rgb]{0.00,0.00,0.50}{##1}}}
\expandafter\def\csname PYG@tok@gu\endcsname{\let\PYG@bf=\textbf\def\PYG@tc##1{\textcolor[rgb]{0.50,0.00,0.50}{##1}}}
\expandafter\def\csname PYG@tok@gd\endcsname{\def\PYG@tc##1{\textcolor[rgb]{0.63,0.00,0.00}{##1}}}
\expandafter\def\csname PYG@tok@gi\endcsname{\def\PYG@tc##1{\textcolor[rgb]{0.00,0.63,0.00}{##1}}}
\expandafter\def\csname PYG@tok@gr\endcsname{\def\PYG@tc##1{\textcolor[rgb]{1.00,0.00,0.00}{##1}}}
\expandafter\def\csname PYG@tok@ge\endcsname{\let\PYG@it=\textit}
\expandafter\def\csname PYG@tok@gs\endcsname{\let\PYG@bf=\textbf}
\expandafter\def\csname PYG@tok@gp\endcsname{\let\PYG@bf=\textbf\def\PYG@tc##1{\textcolor[rgb]{0.00,0.00,0.50}{##1}}}
\expandafter\def\csname PYG@tok@go\endcsname{\def\PYG@tc##1{\textcolor[rgb]{0.53,0.53,0.53}{##1}}}
\expandafter\def\csname PYG@tok@gt\endcsname{\def\PYG@tc##1{\textcolor[rgb]{0.00,0.27,0.87}{##1}}}
\expandafter\def\csname PYG@tok@err\endcsname{\def\PYG@bc##1{\setlength{\fboxsep}{0pt}\fcolorbox[rgb]{1.00,0.00,0.00}{1,1,1}{\strut ##1}}}
\expandafter\def\csname PYG@tok@kc\endcsname{\let\PYG@bf=\textbf\def\PYG@tc##1{\textcolor[rgb]{0.00,0.50,0.00}{##1}}}
\expandafter\def\csname PYG@tok@kd\endcsname{\let\PYG@bf=\textbf\def\PYG@tc##1{\textcolor[rgb]{0.00,0.50,0.00}{##1}}}
\expandafter\def\csname PYG@tok@kn\endcsname{\let\PYG@bf=\textbf\def\PYG@tc##1{\textcolor[rgb]{0.00,0.50,0.00}{##1}}}
\expandafter\def\csname PYG@tok@kr\endcsname{\let\PYG@bf=\textbf\def\PYG@tc##1{\textcolor[rgb]{0.00,0.50,0.00}{##1}}}
\expandafter\def\csname PYG@tok@bp\endcsname{\def\PYG@tc##1{\textcolor[rgb]{0.00,0.50,0.00}{##1}}}
\expandafter\def\csname PYG@tok@fm\endcsname{\def\PYG@tc##1{\textcolor[rgb]{0.00,0.00,1.00}{##1}}}
\expandafter\def\csname PYG@tok@vc\endcsname{\def\PYG@tc##1{\textcolor[rgb]{0.10,0.09,0.49}{##1}}}
\expandafter\def\csname PYG@tok@vg\endcsname{\def\PYG@tc##1{\textcolor[rgb]{0.10,0.09,0.49}{##1}}}
\expandafter\def\csname PYG@tok@vi\endcsname{\def\PYG@tc##1{\textcolor[rgb]{0.10,0.09,0.49}{##1}}}
\expandafter\def\csname PYG@tok@vm\endcsname{\def\PYG@tc##1{\textcolor[rgb]{0.10,0.09,0.49}{##1}}}
\expandafter\def\csname PYG@tok@sa\endcsname{\def\PYG@tc##1{\textcolor[rgb]{0.73,0.13,0.13}{##1}}}
\expandafter\def\csname PYG@tok@sb\endcsname{\def\PYG@tc##1{\textcolor[rgb]{0.73,0.13,0.13}{##1}}}
\expandafter\def\csname PYG@tok@sc\endcsname{\def\PYG@tc##1{\textcolor[rgb]{0.73,0.13,0.13}{##1}}}
\expandafter\def\csname PYG@tok@dl\endcsname{\def\PYG@tc##1{\textcolor[rgb]{0.73,0.13,0.13}{##1}}}
\expandafter\def\csname PYG@tok@s2\endcsname{\def\PYG@tc##1{\textcolor[rgb]{0.73,0.13,0.13}{##1}}}
\expandafter\def\csname PYG@tok@sh\endcsname{\def\PYG@tc##1{\textcolor[rgb]{0.73,0.13,0.13}{##1}}}
\expandafter\def\csname PYG@tok@s1\endcsname{\def\PYG@tc##1{\textcolor[rgb]{0.73,0.13,0.13}{##1}}}
\expandafter\def\csname PYG@tok@mb\endcsname{\def\PYG@tc##1{\textcolor[rgb]{0.40,0.40,0.40}{##1}}}
\expandafter\def\csname PYG@tok@mf\endcsname{\def\PYG@tc##1{\textcolor[rgb]{0.40,0.40,0.40}{##1}}}
\expandafter\def\csname PYG@tok@mh\endcsname{\def\PYG@tc##1{\textcolor[rgb]{0.40,0.40,0.40}{##1}}}
\expandafter\def\csname PYG@tok@mi\endcsname{\def\PYG@tc##1{\textcolor[rgb]{0.40,0.40,0.40}{##1}}}
\expandafter\def\csname PYG@tok@il\endcsname{\def\PYG@tc##1{\textcolor[rgb]{0.40,0.40,0.40}{##1}}}
\expandafter\def\csname PYG@tok@mo\endcsname{\def\PYG@tc##1{\textcolor[rgb]{0.40,0.40,0.40}{##1}}}
\expandafter\def\csname PYG@tok@ch\endcsname{\let\PYG@it=\textit\def\PYG@tc##1{\textcolor[rgb]{0.25,0.50,0.50}{##1}}}
\expandafter\def\csname PYG@tok@cm\endcsname{\let\PYG@it=\textit\def\PYG@tc##1{\textcolor[rgb]{0.25,0.50,0.50}{##1}}}
\expandafter\def\csname PYG@tok@cpf\endcsname{\let\PYG@it=\textit\def\PYG@tc##1{\textcolor[rgb]{0.25,0.50,0.50}{##1}}}
\expandafter\def\csname PYG@tok@c1\endcsname{\let\PYG@it=\textit\def\PYG@tc##1{\textcolor[rgb]{0.25,0.50,0.50}{##1}}}
\expandafter\def\csname PYG@tok@cs\endcsname{\let\PYG@it=\textit\def\PYG@tc##1{\textcolor[rgb]{0.25,0.50,0.50}{##1}}}

\def\PYGZbs{\char`\\}
\def\PYGZus{\char`\_}
\def\PYGZob{\char`\{}
\def\PYGZcb{\char`\}}

\def\PYGZsh{\char`\#}

\def\PYGZhy{\char`\-}
\def\PYGZsq{\char`\'}

\makeatother

\begin{document}
\title{Combinatorial Optimization with Quantum Computers\thanks{This tutorial is a preprint submitted to Engineering Optimization in July 2024. The accepted version can be found at \doi{10.1080/0305215X.2024.2435538}.}}
%
%
\author{Francisco Chicano\inst{1}\orcidID{0000-0003-1259-2990} \and
Gabiel Luque\inst{1}\orcidID{0000-0001-7909-1416} \and
Zakaria Abdelmoiz Dahi\inst{2}\orcidID{0000-0001-8022-4407} \and
Rodrigo Gil-Merino\inst{3}\orcidID{0000-0001-7376-4499}
} 
\authorrunning{F. Chicano, G. Luque, Z. Dahi, and R. Gil-Merino}
%
\institute{ITIS Software, University of Malaga, Spain 
\and
Univ. Lille, Inria, CNRS, Centrale Lille, UMR 9189 CRIStAL, F-59000 Lille, France \and
UNIR, Spain \\
\email{chicano@uma.es, gluque@uma.es, abdelmoiz-zakaria.dahi@inria.fr}
}
\maketitle              
\begin{abstract}
Quantum computers leverage the principles of quantum mechanics to do computation with a potential advantage over classical computers. While a single classical computer transforms one particular binary input into an output after applying one operator to the input, a quantum computer can apply the operator to a superposition of binary strings to provide a superposition of binary outputs, doing computation apparently in parallel. This feature allows quantum computers to speed up the computation compared to classical algorithms. Unsurprisingly, quantum algorithms have been proposed to solve optimization problems in quantum computers. Furthermore, a family of quantum machines called quantum annealers are specially designed to solve optimization problems. In this paper, we provide an introduction to quantum optimization from a practical point of view. We introduce the reader to the use of quantum annealers and quantum gate-based machines to solve optimization problems.

\keywords{Quantum optimization \and quantum computing \and quantum annealer \and quadratic unconstrained binary optimization.}
\end{abstract}
\section{Introduction}
\label{sec:intro}


Quantum physics, as one of modern physics’s fundamental areas, describes behavior with nature at small scales similar to atoms and subatomic levels. Developed in the early part of the 20th century, it is based on aspects that are at odds with classical intuition, such as superposition and entanglement. Superposition means that a quantum system can be in multiple states simultaneously until measured. This contrasts with classical physics, where an object is always in a definite state. Entanglement is a phenomenon in which two or more particles become correlated to one another so that the state of one cannot be independently defined without referencing at least some properties of the other particle(s), even when those separated by large distances~\cite{einstein1935can}. This theoretical framework is currently setting the standard in the field of quantum computing~\cite{feynman1982} and has created the foundation for the development of advanced technologies like lasers and semiconductors.

Using the ideas of quantum physics, quantum computing is a cutting-edge field of computing that can process calculations far more quickly than traditional computers in some situations. Due to quantum superposition, quantum bits, also known as \emph{qubits}, can exist in a simultaneous superposition of both states, in contrast to classical bits, which can only be either 0 or 1. This phenomenon gives qubits a significant processing advantage by enabling them to analyze large amounts of data simultaneously. Moreover, qubits can be coupled by quantum entanglement in a way that makes it irrelevant how far apart they are from one another and, therefore, the state of one qubit instantly influences the state of another~\cite{nielsen2010quantum}.

There are several techniques to physically realize qubits, including superconducting circuits, trapped ions, and quantum dots. Every one of these implementations has advantages and disadvantages of its own. For instance, superconducting qubits, which are used in many of the existing prototypes for quantum computers, are susceptible to noise and decoherence but can also be quickly adjusted. Research into these technological obstacles is ongoing since qubit stability and dependability are crucial to the development of quantum computers in the real world~\cite{clarke2008superconducting}. Even though interest in quantum technology is growing, we are still in the Noisy Intermediate Scale Quantum era (NISQ), where the number of qubits and noise robustness are restricted. 

Multiple quantum computing paradigms exist, each with unique features and uses. The \emph{adiabatic} and \emph{gate-based} models are the two main ones. The most often used paradigm is gate-based, which is based on the application of quantum gates to qubits~\cite{barenco1995elementary}. These gates form \emph{quantum circuits} that allows the implementation of quantum algorithms~\cite{johnston2019programming}. 
Adiabatic quantum computing is based on the \emph{adiabatic theorem}~\cite{born1928beweis}, 
which claims that a quantum system will stay in its ground state if changes to its Hamiltonian (i.e., the mathematical description of the system) happen slowly enough.
By locating a quantum system's lowest energy state, this method is especially useful for tackling optimization problems~\cite{albash2018adiabatic}.

\emph{Optimization} consists in selecting the best option from a large set of options (sometimes with infinity size) and standard algorithms find it difficult to handle the computational complexity of this operation. There are many important optimization problems that are NP-hard, which means that no polynomial time algorithm is known to solve them~\cite{Garey:Johnson1979}. At the same time, optimization is essential to many scientific and industrial applications. 
The potential of quantum optimization algorithms could drive major development in domains like logistics, artificial intelligence, and materials research.
\emph{Quantum annealers} such as those created by D-Wave employ the quantum annealing process in the adiabatic quantum paradigm to solve optimization problems. These machines are very good at solving problems that map to the Ising model and Quadratic Unconstrained Binary Optimization (QUBO). Applications for quantum annealers include scheduling, traffic flow optimization, and even protein folding. These tools are useful for a variety of combinatorial optimization challenges because they can identify ground states of complex Hamiltonians~\cite{willsch2022benchmarking}. 
Quantum gate-based computers can also be used to solve optimization problems. 
The Quantum Approximate Optimization Algorithm (QAOA) and the Variational Quantum Eigensolver (VQE) are two examples of Variational Quantum Algorithms (VQAs) that are specifically designed for the gate-based quantum computers and play a significant role in optimization. These algorithms combine classical optimization techniques with quantum circuits to solve problems more efficiently \cite{bartschi2020grover}.

It is expected that as quantum technology develops, quantum computing will have a greater impact on optimization, enabling breakthroughs that were previously unattainable with classical computing techniques. The relevance of ongoing research and development in quantum computing is underscored by this transformative potential, which can open up new avenues for efficiency and innovation in several industries~\cite{montanaro2016quantum}.

Our goal is to present a thorough and understandable introduction to quantum optimization. In order to use quantum annealers to solve optimization problems, we will guide the reader through the process of transforming different problem types into the QUBO formulation step-by-step. We also demonstrate how gate-based quantum devices can be used to solve these issues. To help our readers get practical experience and refine their topic knowledge, we will provide examples and source code throughout the article to go along with our discussions.

The rest of the paper is organized as follows. In Section~\ref{sec:background}, the fundamental mathematical background required to comprehend the examples in the following sections is introduced. Additionally, it offers a succinct synopsis of the adiabatic and gate-based paradigms' operational principles. Section~\ref{sec:running} outlines the optimization problems that will be used in many of the examples to illustrate the use of quantum computers. Step-by-step instructions for transforming an arbitrary optimization problem into an unconstrained pseudo-Boolean function are given in Section~\ref{sec:opt-qc}. Sections~\ref{sec:opt-annealer} and \ref{sec:opt-gate} show how to optimize the pseudo-Boolean function in a quantum annealer and a gate-based quantum computer, respectively.
Finally, Section~\ref{sec:conclusions} provides a discussion of the research gaps in quantum optimization that should be addressed by future research.

\section{Background}
\label{sec:background}

\subsection{Mathematical Background}
\label{subsec:math-background}

In this section we review several mathematical concepts used along this paper and some notation details.

We will denote with $\Za = \{\ldots, -3, -2, -1, 0, 1, 2, 3, \ldots\}$ the set of integer numbers and with $\Na = \{0, 1, 2, 3, \ldots\}$ the set of natural numbers (non-negative integers). We will use the notation $[n]=\{1,2,\ldots,n\}$ to denote the natural numbers from $1$ to $n$. Let $w \in \Na^n$ be a tuple of $n$ natural numbers, we denote with $|w|=\sum_{j=1}^{n} w_j$ the sum of all the elements in the tuple. We will also use complex number along the paper and we will use $i=\sqrt{-1}$ to denote the imaginary unit. We will avoid the use of $i$ as an index in sums and products, and will use $j$, $k$ and $\ell$ instead.

The matrix exponential of a square matrix $A$ is defined by the following infinite series
\begin{equation}
e^{{A}} = \sum_{k=0}^{\infty} \frac{{A}^k}{k!} ,
\end{equation}
where ${A}^k$ denotes the $k$-th power of matrix ${A}$, and $k!$ is the factorial of $k$.

When the square matrix ${A}$ is diagonalizable, it can be expressed as ${A} = {P} {D} {P}^{-1}$, where 
$P$ is a matrix whose columns are the eigenvectors of $A$, and
${D}$ is a diagonal matrix that contains the corresponding eigenvalues of $A$:
\[
    {D} = \begin{pmatrix}
    \lambda_1 & 0 & \cdots & 0 \\
    0 & \lambda_2 & \cdots & 0 \\
    \vdots & \vdots & \ddots & \vdots \\
    0 & 0 & \cdots & \lambda_n
    \end{pmatrix}.
\]
The matrix exponential can then be computed as
\begin{equation}
    e^{{A}} = {P} e^{{D}} {P}^{-1},
\end{equation}
where $e^{{D}}$ is
\[
    e^{{D}} = \begin{pmatrix}
    e^{\lambda_1} & 0 & \cdots & 0 \\
    0 & e^{\lambda_2} & \cdots & 0 \\
    \vdots & \vdots & \ddots & \vdots \\
    0 & 0 & \cdots & e^{\lambda_n}
    \end{pmatrix} .
\]

When the matrix ${A}$ is not diagonalizable, the exponential can be computed with the help of the Jordan canonical form of the matrix~\cite{Weintraub2009}, but we will not find this situation in the current paper.

We say that two matrices commute if $A B = B A$. We define the \emph{commutator} of $A$ and $B$ as $[A,B]=A B - B A$. If two matrices $A$ and $B$ commute then their commutator is $[A,B]=0$. When $A$ and $B$ commute we also have $e^A e^B = e^{A+B}$. If $A$ and $B$ do not commute the previous expression is not necessarily true.

We will use \emph{Dirac's notation} for denoting vectors $\ket{\psi}$ and covectors $\bra{\psi}$. We will use $\otimes$ to denote the \emph{tensor product} of two vectors and two operators. Since we will only work with the vector space of the computational basis, $\ket{0}$ and $\ket{1}$, we will use string concatenation to denote the tensor product of two vectors. For example, $\ket{00}=\ket{0} \otimes \ket{0}$. Let us assume that we have an operator $A$ acting on a vector space $V_1$ and an operator $B$ acting on a vector space $V_2$, then the tensor product of these two operators, denoted with $A \otimes B$ will act on the tensor product of the vector spaces $V_1 \otimes V_2$ as follows
\begin{equation}
    (A \otimes B) (\ket{\psi} \otimes \ket{\varphi}) = (A \ket{\psi}) \otimes (B \ket{\varphi}) .
\end{equation}

A \emph{Quadratic Unconstrained Binary Optimization} (QUBO) model~\cite{KochenbergerHGLLWW14} is a pseudo-Boolean function (the arguments are binary variables) that can be written as a sum of terms in which there are at most two binary variables multiplied in those terms. Formally,
\begin{eqnarray}
& & f: \{0,1\}^n \rightarrow \mathbb{R}, \\
\label{eqn:qubo-def} & & f(x) = \sum_{j=1}^n \sum_{k=1}^n Q_{j,k} x_j x_k + c,  
\end{eqnarray}
where $Q_{j,k}$ and $c$ are constants and $x_j$ represents the binary variable at the $j$-th position of the binary vector $x$. The constants $Q_{j,k}$ are part of a matrix $Q$, where the value of row $j$ and column $k$ is precisely $Q_{j,k}$. 

In Equation \eqref{eqn:qubo-def} the terms verify the following:
\begin{itemize}
    \item $Q_{j,j} x_j x_j$ are linear terms, since $x_j^2=x_j$ because the variable $x_j$ is binary.
    \item $Q_{j,k}$ and $Q_{k,j}$ are constants that multiply the same product of variables $x_j x_k$. We can write $(Q_{j,k}+Q_{k,j}) x_j x_k$. This gives us an extra degree of freedom for each pair of binary variables that we can take advantage of to simplify some expressions. In what follows, we will assume that the matrix $Q$ is symmetric, i.e.: $Q_{j,k}=Q_{k,j}$.
\end{itemize}

We will also work with higher-order \emph{pseudo-Boolean polynomials}, where the degree of the terms is not restricted to 1 or 2. A useful function related to pseudo-Boolean optimization is the \emph{Iverson bracket}, which takes a Boolean value and return 1 if it is true and 0 if it is false. It will be specially useful to transform a predicate $P$ into a pseudo-Boolean function $[P]$. The Iverson bracket should not be confused with the commutator of two matrices, the latter has two arguments while the Iverson bracket has only one.

An \emph{Ising model} is a function that depends on a set of variables that can take values $-1$ and $1$, originally representing quantum systems of spin $1/2$. The general form is
\begin{eqnarray}
    & & f: \{-1,1\}^n \rightarrow \mathbb{R}, \\
 \label{eqn:ising}   & & f(s) = \sum_{j=1}^n \sum_{k=1}^n J_{j,k} s_j s_k + \sum_{j=1}^n h_j s_j,
\end{eqnarray}
where $J_{j,k}$ and $h_j$ are constants and with $s_j$ we represent the $j$-th variable of the vector $s$. We can assume that $J$ is a matrix and $h$ a vector. All the elements in the diagonal of matrix $J$ will shift the function by a constant because $s_j^2=1$. Thus, we will assume that all the elements $J_{j,j}=0$ except for $J_{1,1}$.

QUBO models and Ising models are equivalent. We can easily transform from one to the other using the variable replacement $x_j=(1-s_j)/2$. We will be specially interested in transforming QUBOs into Ising models. We can do this using the expressions
\begin{align}
\label{eqn:jjk}
    J_{j,k} &= \frac{1}{4} Q_{j,k}, \;\; \text{with $j \neq k$}, \\
\label{eqn:j11}    
    J_{1,1} &= c+\frac{1}{4} \sum_{j,k} Q_{j,k} + \frac{1}{4} \sum_{j=1}^n Q_{j,j}, \\
\label{eqn:jjj}    
    J_{j,j} &= 0 \;\; \text{for $j=2, \ldots, n$}, \\
\label{eqn:hk}    
    h_j &= -\frac{1}{2} \sum_{k=1}^n Q_{j,k},
\end{align}
where we assumed that the $Q$ matrix of the QUBO is symmetric.

Unless stated otherwise, we will assume throughout the article that all objective functions are to be minimized. If we want to maximize an objective function $f$, we only need to multiply $f$ by $-1$ to minimize $-f$.

\subsection{Quantum annealers}
\label{subsec:annealer}

In Section~\ref{sec:intro}, we mentioned that quantum computing can be categorized into two primary paradigms: gate-based and adiabatic. Quantum annealers are a particular implementation of the latter. This section will start describing the adiabatic paradigm, and then explaining how quantum annealers operate as optimization mechanisms.

The \emph{adiabatic theorem} \cite{born1928beweis} is a key notion that forms the basis of adiabatic quantum computing. Essentially, the theorem asserts that a quantum system will stay in its current lowest energy state if the governing Hamiltonian changes at a sufficiently slow rate and if there is a significant energy difference between the lowest energy level and the next highest energy state during the process. If $H(t)$ represents the time-varying Hamiltonian and $\ket{\psi(t)}$ represents the ground state at a certain moment, the system will remain in $\ket{\psi(t)}$ as long as the rate of change of $H(t)$ is significantly slower than the inverse square of the minimum energy gap \cite{jansen2007bounds}. This idea is essential in adiabatic quantum computing because it ensures that the system's ultimate state, after a gradual evolution, will be the lowest energy state of the final Hamiltonian. Consequently, it represents the solution to the computational problem encoded in that Hamiltonian.

Adiabatic quantum computing (AQC) \cite{albash2018adiabatic} leverages the adiabatic theorem to do optimization. It initializes a quantum system in the ground state of a well-known Hamiltonian $H_0$, and then slowly evolves the Hamiltonian to a final one $H_P$ that represents the optimization problem to minimize. According to the adiabatic theorem, at the end of the evolution the system should be in the ground state of the final Hamiltonian, which is the solution to the problem being addressed.

\emph{Quantum annealers} are a specialized form of adiabatic quantum computer that uses quantum annealing optimization process to solve the problem at hand. Quantum annealing \cite{kadowaki1998quantum} is a variant of classical simulated annealing~\cite{kirkpatrick1983optimization}, a stochastic optimization method that simulates the annealing process in metallurgy. In this way, quantum annealing starts by placing the system in a superposition of all possible states. Subsequently, it undergoes a slow conversion into the state of lowest energy in the final Hamiltonian ($H_P$). Quantum tunneling enables the system to move between energy states as the Hamiltonian evolves, thereby surpassing energy barriers that would confine classical systems to local minima.

Therefore, the quantum annealers are mostly distinguished by their adiabatic evolution, which aims to optimize a given function. There are some important differences with respect to AQC. Mainly, quantum annealers start in an initial state that is randomly chosen, usually a uniform superposition, and the system may not initially be in a ground-state and is not required to remain in a ground-state during its evolution.

\subsection{Gate-based quantum computers}
\label{subsec:gate-based}

Now, we focus on the quantum gate-based paradigm. This approach is more similar to classical computing than the adiabatic paradigm because it uses discrete operations, akin to classical logic gates. However, it differs fundamentally in that it exploits quantum superposition and entanglement, enabling potentially exponential speedups for certain problems.

This paradigm describes computations as \emph{quantum circuits} composed of a series of \textit{quantum gates} acting on quantum bits (qubits). A qubit state $\ket{\psi}$ is, in general, a superposition of two base states $\ket{0}$ and $\ket{1}$, where $\ket{\psi} = \alpha \ket{0} + \beta \ket{1}$ and $\alpha, \beta \in \mathbb{C}$. They also verify the normalization condition $|\alpha|^{2} + |\beta|^{2} = 1$. 

The \emph{Bloch sphere} (see Figure~\ref{fig:bloch_sphere}) is a useful geometrical representation of the pure state space of a two-level quantum mechanical system (qubit). Each point on the sphere corresponds to a unique pure state of the qubit. The state $\ket{\psi}$ can be written without loss of generality as
\[
\ket{\psi} = \cos\left(\frac{\theta}{2}\right)\ket{0} + e^{i\varphi}\sin\left(\frac{\theta}{2}\right)\ket{1},
\]
where $\theta$ and $\varphi$ are spherical coordinates. 

\begin{figure}[!ht]
\centering
\begin{tikzpicture}[line cap=round, line join=round, >=Triangle]
  \clip(-2.19,-2.49) rectangle (2.66,2.58);
  
  \draw [shift={(0,0)}, lightgray, fill, fill opacity=0.1] (0,0) -- (56.7:0.4) arc (56.7:90.:0.4) -- cycle;
  \draw [shift={(0,0)}, lightgray, fill, fill opacity=0.1] (0,0) -- (-135.7:0.4) arc (-135.7:-33.2:0.4) -- cycle;
  
  \draw(0,0) circle (2cm);
  \draw [rotate around={0.:(0.,0.)},dash pattern=on 3pt off 3pt] (0,0) ellipse (2cm and 0.9cm);
  
  \draw [thick,->] (0,0) -- (0.7,1.07) node[above right] {$|\psi\rangle$};
  

  \draw [->] (0,0) -- (0,2);
  \draw [->] (0,0) -- (-0.81,-0.79);
  \draw [->] (0,0) -- (2,0);
  
  \draw [dotted] (0.7,1)-- (0.7,-0.46);
  \draw [dotted] (0,0)-- (0.7,-0.46);
  
  \draw (-0.08,-0.3) node[anchor=north west] {$\varphi$};
  \draw (0.01,0.9) node[anchor=north west] {$\theta$};

  \draw (-1.01,-0.72) node[anchor=north west] {$\mathbf {\hat{x}}$};
  \draw (2.07,0.3) node[anchor=north west] {$\mathbf {\hat{y}}$};
  \draw (-0.5,2.6) node[anchor=north west] {$\mathbf {\hat{z}=|0\rangle}$};
  \draw (-0.4,-2) node[anchor=north west] {$-\mathbf {\hat{z}=|1\rangle}$};

  \scriptsize
  \draw [fill] (0,0) circle (1.5pt);
  \draw [fill] (0.7,1.1) circle (0.5pt);
\end{tikzpicture}
\caption{The Bloch sphere representation of a qubit state.}
\label{fig:bloch_sphere}
\end{figure}

The quantum state of a multiple-qubits system can be represented by the tensor product of the qubits' quantum states $\ket{\psi} =a_w \ket{w}$, where $w \in \{0,1\}^n$ and $a_w \in \mathbb{C}$. Thus, each quantum state in a system with $n$ qubits is represented by a tuple of $2^n$ complex numbers. 

Each quantum gate is represented by a \textit{unitary transformation} $U: \mathbb{C}^{2^{n}} \rightarrow \mathbb{C}^{2^{n}}$, which verifies $U^{\dag}U = UU^{\dag} = I$, where $U^{\dag}$ is the Hermitian adjoint of $U$ and $I$ is the identity matrix. Several quantum gates exist~\cite{nielsen2010quantum}. We next describe the most important quantum gates in this paper.

The \emph{Hadamard} gate, denoted as \texttt{H}, is a single-qubit gate that maps the basis states $\ket{0}$ and $\ket{1}$ to equal superpositions of these states. It is represented by the matrix:
\[
H = \frac{1}{\sqrt{2}} \begin{pmatrix}
1 & 1 \\
1 & -1
\end{pmatrix}.
\]
The {Hadamard} gate is used to create superpositions and it is a fundamental gate for many quantum algorithms. For example, when it is applied to a qubit in state $\ket{0}$, it produces the state $\frac{1}{\sqrt{2}}(\ket{0} + \ket{1})$. 

The \emph{Pauli-X} gate, also known as the quantum \texttt{NOT} gate, is analogous to the classical \texttt{NOT} gate and is essential for creating inversions and operations that require state flipping. The \texttt{NOT} gate flips the state of a qubit. It transforms $\ket{0}$ to $\ket{1}$ and $\ket{1}$ to $\ket{0}$. The matrix representation of the \texttt{NOT} gate is
\[
\sigma_{x} = \begin{pmatrix}
0 & 1 \\
1 & 0
\end{pmatrix}.
\]

The \texttt{RX} gate is a rotation around the X-axis of the Bloch sphere. It is represented by the matrix
\[
R_{X}(\theta) = e^{-i \frac{\theta}{2} \sigma_x} = \begin{pmatrix}
\cos(\theta/2) & -i\sin(\theta/2) \\
-i\sin(\theta/2) & \cos(\theta/2)
\end{pmatrix}.
\]


The \emph{Pauli-Y} gate performs a bit flip and a phase flip simultaneously. It is represented by the matrix
\[
\sigma_{y} = \begin{pmatrix}
0 & -i \\
i & 0
\end{pmatrix},
\]
where $i$ is the imaginary unit. This gate transforms $\ket{0}$ to $i\ket{1}$ and $\ket{1}$ to $-i\ket{0}$.
The \texttt{RY} gate is a rotation around the Y-axis of the Bloch sphere. It is represented by the matrix
\[
R_{Y}(\theta) = e^{-i \frac{\theta}{2} \sigma_y} = \begin{pmatrix}
\cos(\theta/2) & -\sin(\theta/2) \\
\sin(\theta/2) & \cos(\theta/2)
\end{pmatrix}.
\]

The \emph{Pauli-Z} gate, also known as the phase-flip gate, changes the sign of the $\ket{1}$ state but leaves the $\ket{0}$ state unchanged. It is characterized by the diagonal matrix
\[
\sigma_{z} = \begin{pmatrix}
1 & 0 \\
0 & -1
\end{pmatrix}.
\]
This gate has a central role in the construction of Hamiltonian representing Ising models. Its operation over the computational basis can be algebraically expressed with the expression $Z\ket{x} = (-1)^{x}\ket{x}$, where $x \in \{0,1\}$.
The \texttt{RZ} gate is a rotation around the Z-axis of the Bloch sphere. It is represented by the matrix
\[
R_{Z}(\theta) =  e^{-i \frac{\theta}{2} \sigma_z} = \begin{pmatrix}
e^{-i\theta/2} & 0 \\
0 & e^{i\theta/2}
\end{pmatrix}.
\]

The Controlled-NOT (\texttt{CNOT}) gate is a two-qubit gate that performs a NOT operation on the target qubit only when the control qubit is in the state $\ket{1}$. Its matrix representation in the computational basis is
\[
\text{CNOT} = \begin{pmatrix}
1 & 0 & 0 & 0 \\
0 & 1 & 0 & 0 \\
0 & 0 & 0 & 1 \\
0 & 0 & 1 & 0
\end{pmatrix}.
\]
The \texttt{CNOT} gate is essential for creating entanglement between qubits, a key resource for quantum computation.

The set composed of the rotation gates \texttt{RX}, \texttt{RY}, \texttt{RZ} plus the \texttt{CNOT} gate is a \emph{universal set of operators}, which means that any unitary transformation can be represented with a quantum circuit containing only these gates.

One important operation (not an operator) in a gate-based quantum computer is the \emph{measurement}.
The measurement operation collapses the quantum state to one of the basis states, providing a classical bit string as the outcome. When a quantum computer in state $\ket{\psi} =\sum_{w \in \{0,1\}^n}a_w \ket{w}$ is measured, the probability of obtaining the binary string $w$ is $|a_w|^2$. The result of the measurement collapses the qubit to the measured state $\ket{w}$. Measurements are crucial for extracting information from quantum computations, as it bridges the quantum and classical worlds.

\section{Running Examples}
\label{sec:running}

In this section we will introduce two NP-hard optimization problems that will serve in the remaining sections to illustrate the steps required to use quantum computers to solve optimization problems. The first optimization problem will be an integer function with constraints. This is a quite general family of problems that will be useful to model how to deal with the constraints in quantum optimization. The second optimization problem will be maximum satisfiability (MAX-SAT)~\cite{Battiti2001}, a well-known optimization problem with no constraints which is at the core of many tools and has a great interest in computer science.



\subsection{Constrained Integer Optimization}

An \textit{integer function} is a function where the input is an integer vector and the output is an integer value, that is, $f: \Za^n \rightarrow \Za$. We will assume in this work that we deal with integer functions that can be written as polynomials of maximum degree $k$ (a constant). This ensures that the maximum number of terms in the polynomial representation of the integer functions is $O(n^k)$. These polynomials can be written in the compact form
\begin{equation}
\label{eqn:pb-polynomial}
p(z) = \sum_{\genfrac{}{}{0pt}{}{w \in  \Za^n}{|w| \leq k}} a_{w} \prod_{j=1}^{n} z_j^{w_j},
\end{equation}
where $z_j$ is the $j$-th integer variable, and the $a_{w}$ constants are the coefficients of the polynomial.

In \textit{constrained integer optimization} the goal is to minimize an integer polynomial $p(z)$ subject to some constraints based on other $m$ integer polynomials $q_j(z)$. In formal terms, we formulate the problem as
\begin{eqnarray}
\label{eqn:constrained-pb-optimization-max}
    & & \min \;\; p(z) \\
    \nonumber \text{subject to:} \\
\label{eqn:constrained-pb-optimization-cons}    
    & & q_j(z) \leq 0 \;\; \forall j \in [m], \\
\label{eqn:constrained-pb-optimization-domain}     
    & & z \in \Za^n .
\end{eqnarray}

For practical reasons, we will need to bound the domain of the integer variables to some interval $[a,b] \subseteq \Za$.

\begin{example}
One problem instance with five variables and six constraints is the following
    \begin{eqnarray*}
        && \min \;\; 5 z_1^3 z_2 - 3 z_2 z_4^3 z_5 + 2 z_3 \\
        \text{subject to:} \\
        && z_4 z_1 + z_2 z_3^2 -1 \leq 0 , \\
        && -5 \leq z_j \leq 5 \;\; \forall j \in [5]
    \end{eqnarray*}
    \qed
\end{example}

\subsection{Maximum Satisfiability (MAX-SAT)}

The MAX-SAT problem~\cite{Battiti2001} is a well-known problem related to the satisfiability of Boolean formulas. 
A \textit{literal} is a Boolean variable $x_j$ or a negated Boolean variable $\neg {x_j}$. 
A positive literal $x_j$ is satisfied when the Boolean variable is true and
a negative literal $\neg x_j$ is satisfied when the Boolean variable is false.
A \textit{clause} is a disjunction of literals (e.g., $x_1 \vee \neg x_2 \vee x_3$), and is satisfied if and only if any of the literals is satisfied.
MAX-SAT instances are composed of a set of clauses $C$. 
The MAX-SAT problem consists in finding an assignment of Boolean values to the variables in such a way that the number of satisfied clauses is maximum or, equivalently, the number of unsatisfied clauses is minimum.
The objective function of MAX-SAT to minimize is defined as
\begin{align}
\label{eqn:f-maxsat}f(x) &= \sum_{c \in C} f_{c}(x), \\
\intertext{where}
f_c(x) &= \left\{\begin{array}{ll}1 & \mbox{if $c$ is unsatisfied with assignment $x$}, \\ 
0 & \mbox{otherwise}.\end{array} \right.
\end{align}

\begin{example}
\label{ex:max-3sat-example}
    One MAX-SAT instance with three variables is the following
    \begin{eqnarray*}
       C= \{ x_1 \vee \neg x_2 \vee x_3 ,
        x_1 \vee x_2 \vee \neg x_3 ,
        \neg x_1 \vee \neg x_2 \vee x_3 \} .
    \end{eqnarray*}
    \qed
\end{example}

\section{Transforming the optimization problem}
\label{sec:opt-qc}

Both quantum annealers and gate-based quantum computers need the optimization problem to be expressed as an unconstrained pseudo-Boolean function. Thus, in this section we will describe how to transform an arbitrary combinatorial optimization problem step by step into a pseudo-Boolean function with no constraints.

\subsection{Solution Representation}
\label{subsec:solution-representation}

The first step in solving an optimization problem using a quantum computer is to decide how to represent the solutions of the problem in the quantum machine.
Most of todays' quantum computers are based on qubits, which can be represented by a 2-dimensional Hilbert space. There are some machines based qudits~\cite{Chicco2024proof} with a higher number of base states (qutrits, ququads, etc.), but we will focus here on the most popular machines, based on qubits. 
The solutions of pseudo-Boolean optimization problems can be naturally represented with the help of qubits and there is nothing to do at this step.
\begin{example}
    In the MAX-SAT problem, each Boolean variable can be mapped into a single qubit of the machine. \qed
\end{example}

\subsubsection{Integer variables}

When we have integer variables we need to express each variable with several Boolean variables. In this case, we need to bound the domain of the integer variable to end with a finite representation of the variable. Let's start with a simple case. Imagine that our integer variable $z \in \Za$ takes values in the interval $[0,2^n-1]$, then we can use $n$ Boolean variables $x_j$ with $j \in [n]$ to represent $z$ as classical computers do
\begin{equation}
\label{eqn:integer-to-boolean-2n}
    z = \sum_{j=0}^{n-1} 2^j x_{j+1}.
\end{equation}
Observe that we number the variables starting in $1$: $x_1$, $x_2$, etc. Replacing $z$ by the expression in $x_j$ given by Equation~\eqref{eqn:integer-to-boolean-2n} in any expression we can remove the integer variable $z$ and add $n$ new Boolean variables.
\begin{example}
    Let $z \in [0,7]$ be an integer variable of an optimization problem. We can introduce three Boolean variables $x_1$, $x_2$, and $x_3$ to represent $z$ as $z=x_1+2x_2+4x_3$. The value $z=5$ is represented by $x_1=1$, $x_2=0$, $x_3=1$.  \qed
\end{example}

If the domain of $z$ is $[0,2^n-1-b]$ with $b > 0$, then Equation~\eqref{eqn:integer-to-boolean-2n} can provide values for $z$ that are out of the domain. In order to avoid this violation of the domain constraint we can reduce the weight of the ``most significant'' bit, $x_n$. We can safely assume that $b < 2^{n-1}$, otherwise, $z$ can be represented with $n-1$ Boolean variables instead of $n$. In this case, we can represent $z$ in general with the expression
\begin{equation}
\label{eqn:integer-to-boolean-truncated}
    z = (2^{n-1}-b)x_{n} + \sum_{j=0}^{n-2} 2^j x_{j+1},
\end{equation}
which is equivalent to Equation~\eqref{eqn:integer-to-boolean-2n} when $b=0$.
\begin{example}
    Let us assume that $z \in [0,5]$. We need $n=3$ variables to represent $z$ and $b=2$. We use the three Boolean variables $x_1$, $x_2$, and $x_3$ and express $z$ as $z=x_1+2x_2+2x_3$. The value $z=5$ is represented by $x_1=1$, $x_2=1$, $x_3=1$, while $z=3$ has two representations: $x_1=1$, $x_2=1$, $x_3=0$; and $x_1=1$, $x_2=0$, $x_3=1$. \qed
\end{example}

The most general domain for an integer variable has the form $[a,c]$. In this case we need $n=\lceil\log_2 (c-a+1)\rceil$ Boolean variables. The interval  can be conveniently re-written as $[a,a+2^n-1-b]$, where $0\leq b < 2^{n-1}$. In this case we can write $z$ in terms of the new $n$ Boolean variables as
\begin{equation}
\label{eqn:integer-to-boolean-general}
    z = a+(2^{n-1}-b)x_{n} + \sum_{j=0}^{n-2} 2^j x_{j+1}.
\end{equation}
\begin{example}
\label{ex:integer-general}
    Let us assume that $z \in [-3,3]$. We need $n=\lceil\log_2 (3-(-3)+1)\rceil=\lceil\log_2 7\rceil=3$ variables to represent $z$. We have $a=-3$ and $b=1$. We use the three Boolean variables $x_1$, $x_2$, and $x_3$ and express $z$ as $z=-3+x_1+2x_2+3x_3$. The value $z=-1$ is represented by $x_1=0$, $x_2=1$, $x_3=0$, while $z=0$ has two representations: $x_1=1$, $x_2=1$, $x_3=0$; and $x_1=0$, $x_2=0$, $x_3=1$. \qed
\end{example}

\subsubsection{Categorical Variables and Permutations}

A variable $y$ is \emph{categorical} when it takes a value from a finite set of options $L=\{\ell_1, \ell_2, \ldots, \ell_k\}$, also called \emph{levels}. We can represent the variable using binary encoding by introducing one binary variable per level $x_\ell$. This variable will be one if and only if the original categorical variable is assigned that level. This form of encoding, called \emph{one-hot encoding}, also requires that the sum of the variables for all the levels is $1$:
\begin{equation}
    \sum_{\ell \in L} x_\ell = 1 .
\end{equation}
\begin{example}
\label{ex:graph-coloring}
    The \emph{Graph Coloring Problem}~\cite{JensenToft2011} consists in assigning one color to each vertex of a graph $G(V,E)$ such that adjacent nodes have different colors. The optimization version of this problem minimizes the number of adjacent nodes with the same color. Assume we are solving an instance of the graph coloring problem where the set of colors is $\{red, blue, green\}$. We can use a categorical variable $y_j$ for each node $j \in V$. If we use one-hot encoding to represent the solutions using binary variables, we can create three variables per node, $x_{j,red}$, $x_{j,blue}$, and $x_{j,green}$ and require
    \begin{equation}
        x_{j,red}+x_{j,blue}+x_{j,green} = 1 \;\; \forall j \in V.
    \end{equation}
    \qed
\end{example}

A \emph{permutation} $\sigma$ of size $n$ is a sequence of numbers from $1$ to $n$ where no number is repeated. We use the notation $\sigma(j)$ to refer to the value at position $j$ in the permutation $\sigma$.
We can also use one-hot encoding to represent permutations.
For each permutation $\sigma$ we can introduce a set of $n^2$ variables $x_{j,k}$ that take value $1$ if and only if $\sigma(j)=k$. Similar to the categorical variables, we require that each position in the permutation takes one value only:
\begin{equation}
\label{eqn:permutations-cons1}
    \sum_{k=1}^{n} x_{j,k} = 1 \;\; \forall j \in [n] .
\end{equation}
The second constraint of permutations is that the numbers cannot be repeated in the sequence:
\begin{equation}
\label{eqn:permutations-cons2}
    \sum_{j=1}^{n} x_{j,k} = 1  \;\; \forall k \in [n] .
\end{equation}

\begin{example}
    The famous \emph{Traveling Salesperson Problem} (TSP)~\cite{Applegate2006} consists in finding an order to visit a set of $n$ cities such that each city is visited once and only once and the total length of the tour is minimized. A solution to this problem is usually represented by a permutation $\sigma$. We can introduce binary variables $x_{j,k}$ constrained by Equations~\eqref{eqn:permutations-cons1} and~\eqref{eqn:permutations-cons2} to represent the solutions of the problem. \qed
\end{example}

\subsection{Objective Function}
\label{subsec:objetive-function}

Once we have introduced the binary variables required to represent our solutions we need to replace the original variables by the binary ones in the objective function. If the original variables were binary we only need to express the objective functions as a pseudo-Boolean expression. Let's illustrate this with the MAX-SAT instance in Example~\ref{ex:max-3sat-example}.

\begin{example}
\label{ex:maxsat-objective-function}
    A logic clause of a MAX-SAT instance is unsatisfied only if all the literals are false, and satisfied otherwise. Let's consider the first clause of the instance in Example~\ref{ex:max-3sat-example}: $x_1 \vee \neg x_2 \vee x_3$. This clause is unsatisfied when $x_1=0$, $x_2=1$, and $x_3=0$. We can build a cubic polynomial that is $1$ when the clause is unsatisfied: $(1-x_1) x_2 (1-x_3)$. The expression is a product in which the negated variables appear as is and the positive literals appear complemented. The objective function for MAX-SAT is the sum of these expression for all the clauses. In our example,
    \begin{align}
        \nonumber f(x) &= (1-x_1) x_2 (1-x_3) \\
        \nonumber &+ (1-x_1) (1-x_2) x_3 \\
        \nonumber &+ x_1 x_2 (1-x_3) \\
     \label{eqn:maxsat-objective}   &= x_1 x_2 x_3 - 2 x_2 x_3-x_1 x_3+x_2+x_3 .
    \end{align}
    \qed 
\end{example}

If the original function has integer variables, we replace the integer variables by the corresponding binary expressions and expand the expression taking into account that for each binary variable $x$ we have $x^k=x$ for $k \geq 1$. This means that we can simplify the expressions to reach multilinear monomials (terms that are linear in all the variables).
\begin{example}
\label{ex:integer-function}
    Assume that our objective function is $f(z)=z^3-6z$ with $z$ integer and $z \in [-3,3]$. According to Example~\ref{ex:integer-general} we should introduce three binary variables to represent $z$ and express it as $z=-3+x_1+2x_2+3x_3$. Applying this variable change in the objective function we get
    \begin{align*}
        f(x)&=\left(-3+x_1+2x_2+3x_3\right)^3 - 6 \left(-3+x_1+2x_2+3x_3\right) \\
        &= x_1^3+6 x_2 x_1^2+9 x_3 x_1^2-9 x_1^2+12 x_2^2 x_1+27 x_3^2 x_1-36 x_2 x_1+36 x_2 x_3 x_1\\
        &-54 x_3 x_1+21 x_1+8 x_2^3+27 x_3^3-36 x_2^2+54 x_2 x_3^2-81 x_3^2+42 x_2+36 x_2^2 x_3\\
        &-108 x_2 x_3+63 x_3-9\\
        \intertext{using $x^k=x$ for $k \geq 1$}
        &= x_1+6 x_2 x_1+9 x_3 x_1-9 x_1+12 x_2 x_1+27 x_3 x_1-36 x_2 x_1+36 x_2 x_3 x_1\\
        &-54 x_3 x_1+21 x_1+8 x_2+27 x_3-36 x_2+54 x_2 x_3-81 x_3+42 x_2+36 x_2 x_3\\
        &-108 x_2 x_3+63 x_3-9\\
        &=-9+13 x_1+14 x_2+9 x_3-18 x_1 x_2-18 x_1 x_3-18 x_2 x_3+36 x_1 x_2 x_3
    \end{align*}
    \qed
\end{example}

When the problem has categorical variables, a very general way to express the objective function is in terms of other subfunctions based on predicates including the categorical variables. These subfunctions can be written as
\begin{equation}
\label{eqn:categorical-subfunction}
    f(u, v) = \left\{\begin{array}{ll}
    \alpha & \text{if $P(u,v)$ is true,} \\
    \beta & \text{otherwise,}
    \end{array}\right.
\end{equation}
where $\alpha$, $\beta$ are real values, and $P(u,v)$ is a predicate of the categorical variables $u$ and $v$. We can write Equation~\eqref{eqn:categorical-subfunction} in a more compact way as $f(u,v)=(\alpha-\beta)[P(u,v)]+\beta$ using the Iverson bracket. In order to, replace the categorical variables by the binary variables we only need to provide expressions for the Iverson bracket. These expression will depend on the structure of the predicate. Table~\ref{tab:iverson} provides a recursive definition of the Iverson bracket where the base case is always depending on the binary variables.

\begin{table}[!ht]
\caption{Recursive definition of the Iverson bracket for predicates with categorical variables. In the last two rows $u$ and $v$ are two categorical variables with the same levels. The $x$ binary variables are associated to categorical variable $u$ and the $y$ binary variables are associated to categorical variable $v$. }
    \label{tab:iverson}
    \centering
    \begin{tabular}{@{\phantom{0}}l@{\phantom{000}}l@{\phantom{0}}}
        \toprule
        Predicate & Function \\
         \midrule
        $[P \vee Q]$ &  $[P] + [Q] - [P][Q]$\\
        $[P \wedge Q]$ &  $[P][Q]$ \\
        $[\neg P]$ &  $1-[P]$ \\
        $[u=\ell]$ &  $x_\ell$ \\
        $[v=u]$ &  $\sum_{\ell \in L} x_{\ell} y_{\ell}$ \\
         \bottomrule
    \end{tabular}
\end{table}

\begin{example}
    Let us write the objective function for the graph coloring problem, defined in Example~\ref{ex:graph-coloring}. For each edge $(j,k) \in E$ in the graph we need a subfunction that takes value $1$ if the two adjacent nodes have the same color and $0$ if they have different colors. Let $u_j$ be a categorical variable to denote the color of node~$j$. The original objective function is
    \begin{equation}
        f(u) = \sum_{(j,k) \in E} [u_j=u_k],
    \end{equation}
and using the binary variables we can write
    \begin{equation}
        f(x) = \sum_{(j,k) \in E} \sum_{\ell \in L} x_{j,\ell}x_{k,\ell},
    \end{equation}
where $L$ is the set of possible colors. As we explained in Example~\ref{ex:graph-coloring}, we also need to add an equation associated to the constraint that each node has one and only one color: $\sum_{\ell \in L} x_{j,\ell} = 1$ for all $j \in V$. \qed
\end{example}

We can build the objective function of a permutation problem using also the Iverson bracket. If $P(\sigma)$ is a predicate based on permutation $\sigma$, we can use subfunctions with form $f(\sigma)=(\alpha-\beta)[P(\sigma)]+\beta$ to write the objective function. A very common permutation predicate is $\sigma(i)=j$. We can find this predicate in problems like the \emph{Traveling Salesperson Problem} (TSP)~\cite{Applegate2006} and the \emph{Quadratic Assignment Problem} (QAP)~\cite{Cela1998}.

\begin{example}
    \label{ex:linear-assignment}
    The objective function of the \emph{Linear Assignment Problem} can be written as
    \begin{equation}
        f(\sigma) = \sum_{j=1}^{n} c_{j,\sigma(j)} \; .
    \end{equation}
    In terms of the Iverson bracket we can write
    \begin{equation}
        f(\sigma) = \sum_{j=1}^{n}\sum_{k=1}^{n} c_{j,k}[\sigma(j)=k] \;.
    \end{equation}
    Now we can take into account that binary variable $x_{j,k}$ is 1 if and only if $\sigma(j)=k$, by definition. This means $x_{j,k}=[\sigma(j)=k]$ and we finally write the objective function as the pseudo-Boolean expression
    \begin{equation}
        f(x) = \sum_{j=1}^{n}\sum_{k=1}^{n} c_{j,k}x_{j,k} \;,
    \end{equation}
    where we need to add the constraints associated to permutations: Equations~\eqref{eqn:permutations-cons1} and~\eqref{eqn:permutations-cons2}. \qed
\end{example}

\subsection{Constraints}
\label{subsec:constraints}

The pseudo-Boolean functions that a quantum computer can optimize are unconstrained. 
Our next step in the transformation of our optimization problem consists in including the problem constraints in our final objective function. We say that a solution to the problem is \emph{feasible} when it meets all the constraints and \emph{infeasible} when it does not fulfill at least one constraint. In order to combine the constraints and the objective function in one single expression to minimize we introduce \emph{penalty expressions} that increase the value objective function if the solutions are \emph{infeasible}. We illustrate this step in Figure~\ref{fig:constraints}.

\begin{figure}[!ht]
\resizebox{0.9\textwidth}{!}{
\begin{tikzpicture}

    \draw[->] (0,0) -- (14,0);
    \draw[->] (0,0) -- (0,8);
    \node[rotate=90,anchor=south] at (0,3) {Objective value};

    \fill[blue!30] (1,1) rectangle (3,5);
    \draw[blue,thick] (0.9,1) -- (3.1,1);
    \draw[blue,thick] (0.9,5) -- (3.1,5);
    \node[text width=1.5cm,align=center] at (2,3) {All solutions};

    \fill[darkgreen!30] (4,2) rectangle (6,4);
    \draw[darkgreen,thick] (3.9,4) -- (6.1,4);
    \draw[darkgreen,thick] (3.9,2) -- (6.1,2);
    \node[text width=1.5cm,align=center] at (5,3) {Feasible solutions};
    
    \fill[red!30] (7,1) rectangle (9,5);
    \draw[darkred,thick] (6.9,1) -- (9.1,1);
    \draw[darkred,thick] (6.9,5) -- (9.1,5);
    \node[text width=1.5cm,align=center] at (8,3) {Infeasible solutions};

    \draw[darkred,dashed] (9,1) -- (12,1);
    \draw[->] (11,1) -- (11,3);
    \node[anchor=west] at (11,2) {Penalty};

    \fill[red!30] (10,3) rectangle (12,7);
    \draw[darkred,thick] (9.9,3) -- (12.1,3);
    \draw[darkred,thick] (9.9,7) -- (12.1,7);
    \node[text width=1.5cm,align=center] at (11,5) {Penalized infeasible solutions};

    \node[thick] at (3.5,3) {\huge $=$};
    \node[thick] at (6.5,3) {\huge $+$};
\end{tikzpicture}
}

\caption{The solutions in our problem can be classified as feasible and infeasible. In order to build one single expression for our objective function we need to ensure that any infeasible solution has an objective value that is greater than the best feasible solution. We do this by adding penalty expressions to the objective function.}
\label{fig:constraints}
\end{figure}

The expressions in the constraints of an optimization problem must be transformed following the same steps given in Sections~\ref{subsec:solution-representation} and~\ref{subsec:objetive-function}. Let us focus on one constraint and let us assume that it is an equality $h(x)=0$. We also assume that $h(x)$ is integer-valued.\footnote{If there are decimal numbers in the coefficients of $h(x)$, we can multiply the whole function by a large enough integer number and have only integers in the coefficients.} Then, $h^2(x)$ is an appropriate penalty expression that will be equals to 0 if the constraint is satisfied and greater than zero otherwise. We usually need to multiply $h^2(x)$ by a penalty constant $P$ in order to avoid any overlap between infeasible solutions and the best feasible solutions (Figure~\ref{fig:constraints}). We will detail later in this section how this constant $P$ can be estimated.

If the constraint is an inequality $h(x) \leq 0$, then we transform it into an equality constraint using a slack variable. That is, we introduce positive integer variable $s \in \Za$ with $s \geq 0$ and replace the inequality by the equality constraint $h(x)+s=0$, which is transformed into the penalty expression $(h(x)+s)^2$, where $s$ needs to be replaced by the appropriate binary variables. The maximum value of $s$ will be equals to $-\min_{x \in \{0,1\}^n} h(x)$. This value will help us to determine the binary representation of $s$ following the steps in Section~\ref{subsec:solution-representation}.

\begin{example}
    If we add the constraint $3x_1 x_4 x_6 - 4 x_3 x_5 \leq 0$ to our optimization problem, a possible penalty expression is $(3x_1 x_4 x_6 - 4 x_3 x_5 + s)^2$. It is easy to see that $\min_{x \in \Bo^n} (3x_1 x_4 x_6 - 4 x_3 x_5) = -4$, and, thus, the slack variable $s$ should be $0\leq s \leq 4$. We need three new binary variable to represent $s$, let's call them $t_1$, $t_2$ and $t_3$. Variable $s$ can be replaced by $t_1+2t_2+t_3$ and the final penalty expression is
    \begin{equation}
        P (3x_1 x_4 x_6 - 4 x_3 x_5 + t_1+2t_2+t_3)^2,
    \end{equation}
    where $P$ will depend on the objective function of the problem. \qed
\end{example}

Regarding the penalty constant $P$, the interested reader can find research work devoted only to determine the values for this constant~\cite{DBLP:conf/evoW/Ayodele22}. We will provide here a rule to find a value that works, but not necessarily the best value to solve an optimization problem. The idea in this rule is to focus on the objective function $f(x)$, which at this step is a pseudo-Boolean polynomial
\begin{equation}
\label{eqn:pseudo-Boolean}
    f(x) = \sum_{S \subseteq [n]} a_S \prod_{j \in S} x_j .
\end{equation}

Now we claim that a penalty constant $P$ that works is the difference between an upper bound and a lower bound of $f$ plus one: $P=ub(f)-lb(f)+1$. The rationale is as follows. We know an optimal (feasible) solution to our optimization problem, denoted with $x^*$, will be between the lower and upper bound: $lb(f) \leq f(x^*) \leq ub(f)$. Let us denote with $y$ an infeasible solution. We also have $lb(f) \leq f(y) \leq ub(f)$, but we don't know the relative order between $f(y)$ and $f(x^*)$. In particular, it could happen that $f(y) < f(x^*)$, and $y$ is better for our minimization problem than $x^*$. If we add $P=ub(f)-lb(f)+1$ to $f(y)$ we can be sure that $f(y)+P > ub(f)$ as we prove next:
\begin{align*}
    f(y) + P &\geq lb(f) + ub(f) - lb(f) +1 = ub(f)+1 > ub(f).
\end{align*}
Since $f(x^*) \leq ub(f)$, we can ensure that no infeasible solution can be a global optimal solution of the penalized expression.

We only need to compute lower and upper bounds for $f$ to complete the penalty constant computation. We can easily find upper and lower bounds for $f(x)$ based on Equation~\eqref{eqn:pseudo-Boolean}. In order to find a lower bound, we can pessimistically assume that there is a solution $x \in \Bo^n$ such that any non-constant monomial $a_S \prod_{j \in S} x_j$ with $a_S < 0$ takes value $a_S$ and any non-constant monomial with $a_S > 0$ is zero. This way, a lower bound is
\begin{equation}
    lb (f) = a_{\emptyset} + \sum_{S \subseteq [n] \atop S \neq \emptyset} \min(0,a_S).
\end{equation}
A similar approach to compute an upper bound yields
\begin{equation}
    ub (f) = a_{\emptyset} + \sum_{S \subseteq [n] \atop S \neq \emptyset} \max(0,a_S).
\end{equation}
We can now compute the penalty $P$ as
\begin{equation}
\label{eqn:penalty-constant}
    P = ub(f)-lb(f)+1 = 1+ \sum_{S \subseteq [n] \atop S \neq \emptyset} \left(\max(0,a_S)-\min(0,a_S)\right) = 1+ \sum_{S \subseteq [n] \atop S \neq \emptyset} |a_S|.
\end{equation}

\begin{example}
    Let's add the constraint $(z-1)^2\leq 6$ to the integer function in Example~\ref{ex:integer-function}. The constraint function is $h(z)=(z-1)^2-6$. After replacing $z$ by $-3 + x_1 + 2x_2 + 3x_3$ and simplifying we get the expression
    \begin{equation}
        h(x) = 4 x_2 x_1+6 x_3 x_1-7 x_1-12 x_2+12 x_2 x_3-15 x_3+10 .
    \end{equation}
    Instead of computing $\min h(x)$, we compute a lower bound of $h$ using the same procedure described above for the objective function $f$. We get $lb(h)=-7-12-15+10=-24$. The slack variable $s$ in the new equation $h(x)+s=0$, should take values in $[0,24]$. We need five new binary variables to replace slack variable $s$, which we will denote with $t_1$ to $t_5$. The expression for the slack variable is $s=t_1+2t_2+4t_3+8t_4+9t_5$. The penalty constant $P$ can be computed with Equation~\eqref{eqn:penalty-constant}:
    \begin{equation}
        P = 1+ \sum_{S \subseteq [n] \atop S \neq \emptyset} |a_S| = 1+ |13| + |14| + |9| + |-18| + |-18| + |-18| + |36| = 127.
    \end{equation}
    We can finally write the expression to maximize including the penalty expression as $f'(x,t) = f(x) + P(h(x)+s)^2$.
    \qed
\end{example}

\section{Solving the problem in the Quantum Annealer}
\label{sec:opt-annealer}

Quantum annealers (see Section~\ref{subsec:annealer}) are able to find the state of minimum energy of a Hamiltonian that contains linear terms depending on the qubits and pairwise interactions among them. In order to use a quantum annealer we need to express our objective function as an Ising model with pairwise interactions. It is usually easier to first transform the original problem to QUBO and then transform the QUBO into an Ising model using the formulas we saw in Section~\ref{subsec:math-background}. We have seen in Section~\ref{sec:opt-qc} how to transform our optimization problem to an unconstrained pseudo-Boolean polynomial. In this section we will describe how to get a QUBO from the pseudo-Boolean polynomial and how to solve the QUBO in a quantum annealer.

\subsection{Order Reduction}
\label{subsec:order-reduction}

Quantum annealers require the pseudo-Boolean polynomial to be quadratic. 
The quadratic and linear terms already existing in the polynomial do not need to be changed. We will focus on higher-order terms. We can identify in the literature two strategies to reduce the order of a pseudo-Boolean polynomial. The first strategy consists in replacing products of two variables $x y$ by a new variable $z$. If we do this in any term containing the variables $x$ and $y$, we reduce the order of those terms by one. In order to keep the value of $z$ equals to the product of $x$ and $y$ we introduce a penalty expression due to Rosenberg~\cite{Rosenberg1975}: $P(xy - 2xz -2yx + 3z)$. The reader can easily check that when $z=xy$ the expression is zero, but if $z \neq xy$ the expression is $P$ (when $xy=1$ and $z=0$) or $3P$ (when $xy=0$ and $z=1$). This penalty expression is added to the polynomial as any other problem constraint. Applying this procedure iteratively to each monomial with three or more variables we will end with a quadratic expression. This kind of transformation is called \emph{global} because each appearance of the product of variables can be replaced by the new variable and a global penalty term is added to the final pseudo-Boolean polynomial.

\begin{example}
\label{ex:maxsat-rosenberg}
    Let's consider the MAX-SAT instance in Example~\ref{ex:max-3sat-example}. The objective function, computed in Example~\ref{ex:maxsat-objective-function} is $x_1 x_2 x_3 - 2 x_2 x_3-x_1 x_3+x_2+x_3$. There is a cubic term in that expression that we will remove with the help of Rosenberg's penalty expression. We introduce variable $z=x_2x_3$. The final objective function is
    \begin{equation}
    \label{eqn:maxsat-qubo}
        f(x,z) =  x_1 z - 2 z-x_1 x_3+x_2+x_3 +P (x_2 x_3 - 2x_2 z -2 x_3 z + 3z),
    \end{equation}
    where the penalty constant $P$ can be computed using the expression provided in Equation~\eqref{eqn:penalty-constant}. Observe that Equation~\eqref{eqn:maxsat-qubo} is a QUBO. \qed    
\end{example}

The second strategy to reduce the order consists in applying a local transformation to each high-order monomial. This transformation replace the monomial by a quadratic expression with the same optimal solution. One example of these local transformations in minimization is replacing a monomial with negative coefficient $a \prod_{j\in S} x_j$, where $a < 0$ by the quadratic expression $a(\sum_{j \in S} x_j - |S|+1)t$, where $t$ is a new variable. If there are several high order monomials, each monomial will require a new variable $t$; this is why these transformation are called \emph{local}. While this approach can require many new variables, the advantage is that we do not have to provide a penalty constant $P$.
\begin{example}
\label{ex:local-order-reduction}
    Let's imagine that we want to maximize the objective function of Example~\ref{ex:integer-function}. We can transform the problem to minimization by multiplying the objective function by $-1$. The only cubic term in the expression for $-f(x)$ is $-36 x_1 x_2 x_3$, which has a negative coefficient. Thus, we can apply the local transformation $-36 x_1 x_2 x_3 \rightarrow -36 (x_1 + x_2 + x_3 - 2) t$. The final objective function to minimize is 
    \begin{equation}
        f'(x,t) = 9 - 13x_1 - 14x_2 - 9x_3 + 18x_1x_2 + 18x_1x_3 + 18x_2x_3 - 36 (x_2 + x_2 + x_3 - 2) t,
    \end{equation}
    where $t$ is a new binary variable. \qed
\end{example}

We only showed one example of local transformation that is valid when the coefficient of the monomial is positive in negative in minimization (or positive in maximization). The reader interested in local transformations is referred to the compilation done by Nike Dattani~\cite{dattani2019quadratization}.

\subsection{Using the quantum annealer}

Quantum annealers are specialized in minimizing Ising models. Once we have a QUBO, we can transform it into an Ising model using Equations~\eqref{eqn:jjk} to~\eqref{eqn:hk}.
In the final Ising model, we can omit the constant term, since it only shifts the objective function but do not change the global optimal solution. Thus, we can do $J_{j,j}=0$ for all $j$. When two variables $s_j$ and $s_k$ are mapped to two qubits in the quantum annealer, the coefficient $J_{j,k}$ is called \emph{coupling} between the two qubits.

\begin{figure}[!ht]
    \centering
    \begin{subfigure}{0.3\textwidth}
        \centering
        \includegraphics[width=\textwidth]{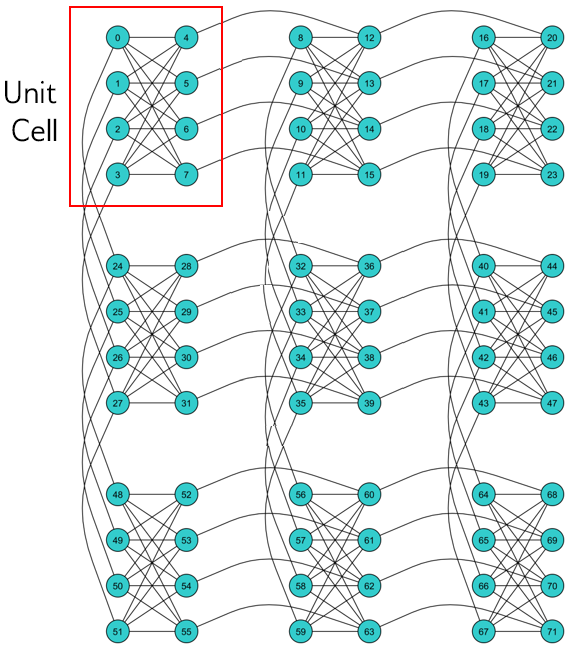}
        \caption{Chimera}
        \label{dwave-arch-chimera}
    \end{subfigure}
    \hfill
    \begin{subfigure}{0.3\textwidth}
        \centering
        \includegraphics[width=\textwidth]{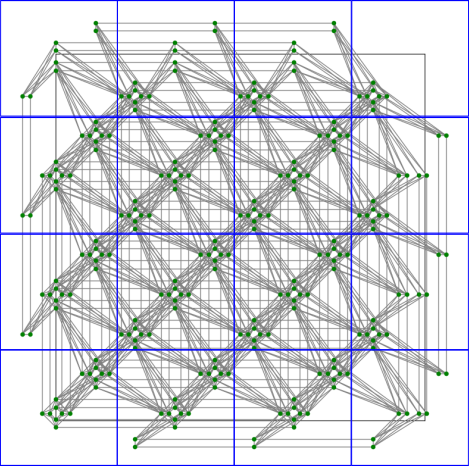}
        \caption{Pegasus}
        \label{dwave-arch-pegasus}
    \end{subfigure}
    \hfill
    \begin{subfigure}{0.3\textwidth}
        \centering
        \includegraphics[width=\textwidth]{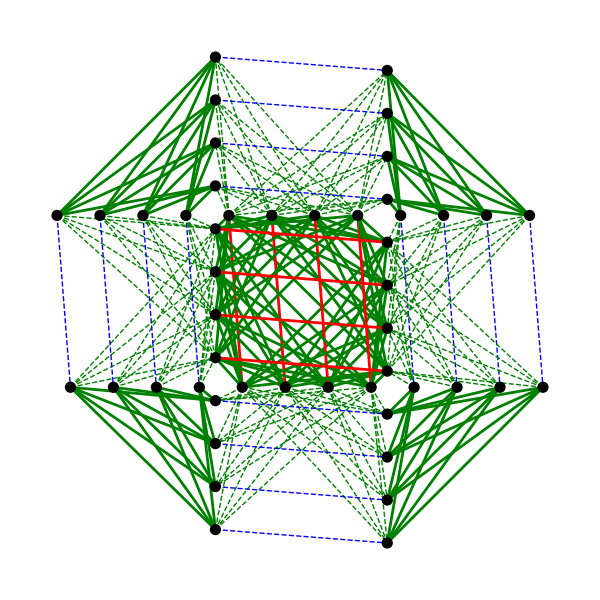}
        \caption{Zephyr}
        \label{dwave-arch-zephyr}
    \end{subfigure}
    \caption{Graph topology of D-Wave machines. Source: D-Wave official documentation (reproduced with permission).}
    \label{fig:dwave-arch}
\end{figure}

Modern quantum annealers do not allow to have a coupling among any arbitrary pair of qubits, the interaction is restricted to some pairs that are determined by the topology of the quantum annealer. In Figure~\ref{fig:dwave-arch} we show three different connection topologies of D-Wave machines: \textit{Chimera}, \textit{Pegasus} and \textit{Zephyr}. In practice, the constraint imposed by the topology means that we cannot set a non-zero value for any arbitrary $J_{j,k}$ coefficient in the Ising model. In order to solve this issue, instead of mapping one variable $s_j$ to a single qubit, we need to represent one variable with a set of connected qubits in the quantum annealer, also called a \emph{chain}. The coupling between the qubits in the chain must be negative and high in absolute value in order to avoid that the different qubits of the chain take different values. When there is a qubit in a chain that has a value different to the others in the final quantum state, the variable represented by the chain does not have a defined value in that quantum state. The problem of mapping the Ising model to the particular topology of a quantum machine is called \emph{minor-embedding}, which is NP-hard~\citep{Liu2021}.

Current software tools to use quantum annealers help the user to do some of the previous steps automatically. For example, it is usually not required to transform the QUBO into an Ising model. It is also not required to find constants for the penalty expressions, because they compute the appropriate values from the constraints themselves. On prominent example of Open Source SDK for optimizing combinatorial optimization problems using quantum annealers is D-Wave Ocean\footnote{\url{https://www.dwavesys.com/solutions-and-products/ocean/}}. Figure~\ref{fig:qubo-dwave} shows a simple code in Python using D-Wave Ocean SDK to solve the MAX-SAT instance of Example~\ref{ex:max-3sat-example} using the fitness function computed in Example~\ref{ex:maxsat-rosenberg} with penalty constant $P=7$. The results obtained running the code in the D-Wave's Advantage 4.1 system is shown in Table~\ref{tab:results-dwave}. Each row represents a different combination of the binary variables at the end of the annealing process that has been obtained in the 1000 runs. We can observe that there are three assignments that satisfy all the clauses and they appear more frequently that the other combinations.

\begin{figure}[!ht]
\begin{Verbatim}[commandchars=\\\{\},numbers=left,firstnumber=1,stepnumber=1,fontsize=\scriptsize,frame=lines]
\PYG{k+kn}{from} \PYG{n+nn}{dimod} \PYG{k+kn}{import} \PYG{n}{Binary}\PYG{p}{,} \PYG{n}{ExactSolver}
\PYG{k+kn}{from} \PYG{n+nn}{dwave.system} \PYG{k+kn}{import} \PYG{n}{DWaveSampler}\PYG{p}{,} \PYG{n}{EmbeddingComposite}
\PYG{n}{qubo} \PYG{o}{=} \PYG{n}{Binary}\PYG{p}{(}\PYG{l+s+s1}{\PYGZsq{}x1\PYGZsq{}}\PYG{p}{)} \PYG{o}{*} \PYG{n}{Binary}\PYG{p}{(}\PYG{l+s+s1}{\PYGZsq{}z\PYGZsq{}}\PYG{p}{)} \PYG{o}{\PYGZhy{}} \PYG{l+m+mi}{2} \PYG{o}{*} \PYG{n}{Binary}\PYG{p}{(}\PYG{l+s+s1}{\PYGZsq{}z\PYGZsq{}}\PYG{p}{)} \PYGZbs{}
  \PYG{o}{\PYGZhy{}} \PYG{n}{Binary}\PYG{p}{(}\PYG{l+s+s1}{\PYGZsq{}x1\PYGZsq{}}\PYG{p}{)} \PYG{o}{*} \PYG{n}{Binary}\PYG{p}{(}\PYG{l+s+s1}{\PYGZsq{}x3\PYGZsq{}}\PYG{p}{)} \PYG{o}{+} \PYG{n}{Binary}\PYG{p}{(}\PYG{l+s+s1}{\PYGZsq{}x2\PYGZsq{}}\PYG{p}{)} \PYG{o}{+} \PYG{n}{Binary}\PYG{p}{(}\PYG{l+s+s1}{\PYGZsq{}x3\PYGZsq{}}\PYG{p}{)} \PYGZbs{}
  \PYG{o}{+} \PYG{l+m+mi}{7}\PYG{o}{*}\PYG{p}{(}\PYG{n}{Binary}\PYG{p}{(}\PYG{l+s+s1}{\PYGZsq{}x2\PYGZsq{}}\PYG{p}{)}\PYG{o}{*}\PYG{n}{Binary}\PYG{p}{(}\PYG{l+s+s1}{\PYGZsq{}x3\PYGZsq{}}\PYG{p}{)} \PYG{o}{\PYGZhy{}} \PYG{l+m+mi}{2} \PYG{o}{*} \PYG{n}{Binary}\PYG{p}{(}\PYG{l+s+s1}{\PYGZsq{}x2\PYGZsq{}}\PYG{p}{)} \PYG{o}{*} \PYG{n}{Binary}\PYG{p}{(}\PYG{l+s+s1}{\PYGZsq{}z\PYGZsq{}}\PYG{p}{)} \PYG{o}{+} \PYG{l+m+mi}{3} \PYG{o}{*} \PYG{n}{Binary}\PYG{p}{(}\PYG{l+s+s1}{\PYGZsq{}z\PYGZsq{}}\PYG{p}{))}
\PYG{n}{sampler} \PYG{o}{=} \PYG{n}{EmbeddingComposite}\PYG{p}{(}\PYG{n}{DWaveSampler}\PYG{p}{())}
\PYG{n}{result}\PYG{o}{=}\PYG{n}{sampler}\PYG{o}{.}\PYG{n}{sample}\PYG{p}{(}\PYG{n}{qubo}\PYG{p}{,} \PYG{n}{num\PYGZus{}reads}\PYG{o}{=}\PYG{l+m+mi}{1000}\PYG{p}{)}
\PYG{k}{print}\PYG{p}{(}\PYG{n}{result}\PYG{p}{)}
\end{Verbatim}
\caption{Code using D-Wave Ocean SDK to solve Example~\ref{ex:maxsat-rosenberg}.}
\label{fig:qubo-dwave}
\end{figure}

\begin{table}[!ht]
\centering
\caption{Results obtained after running the code in Figure~\ref{fig:qubo-dwave}. The energy is the value of the objective function (unsatisfied clauses) for each combination of variables and the last column shows the number of times the combination in the row has been seen in the runs.}
\label{tab:results-dwave}
\begin{tabular}{rrrrrr}
\toprule
$x_1$ & $x_2$ & $x_3$ & $z$ & Energy & Occurrences \\
\midrule
1 & 0 & 0 & 0 & 0.0 & 371 \\
1 & 0 & 1 & 0 & 0.0 & 320 \\
0 & 0 & 0 & 0 & 0.0 & 270 \\
1 & 1 & 0 & 0 & 1.0 &  16 \\
0 & 0 & 1 & 0 & 1.0 &  15 \\
0 & 1 & 0 & 0 & 1.0 &   8 \\
\bottomrule
\end{tabular}
\end{table}

\section{Optimization in a Gate-based Machine}
\label{sec:opt-gate}

\begin{figure}[h!]    
\centering
    \includegraphics[scale=0.4]{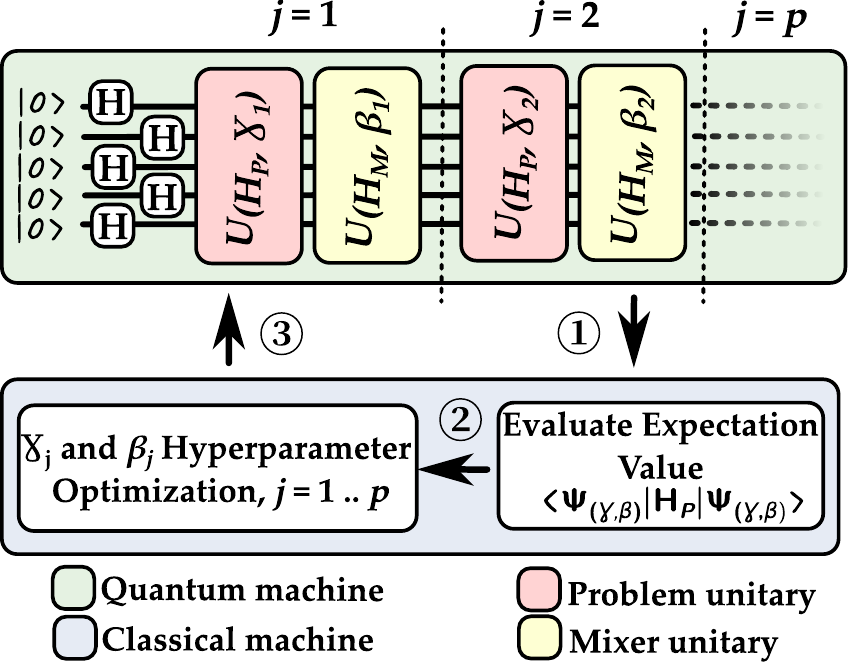}
    \caption{Workflow of QAOA}
    \label{fig:QAOA}
\end{figure}

The Quantum Approximate Optimization Algorithm (QAOA) is one of the recently proposed and investigated quantum optimization techniques in gate-based quantum computers \cite{ref_1}. It is a hybrid quantum algorithm composed of both a quantum and classical parts. The quantum part is composed of a parameterized circuit that is used to sample the solutions to the problem being solved, while classical part is responsible for optimizing the parameters of the quantum part so as to increase the probability of sampling the best solutions. When going further into details, the QAOA quantum part is a type of variational quantum circuit, also called \emph{ansatz}, composed of a given number $p$ of layers, where each layer is described by the product of two unitary transformations $U(H_{P}, \gamma_k)=e^{-i \gamma_{k} {H}_{P}}$ and $U(H_{M}, \beta_k)=e^{-i \beta_{k} {H}_{M}}$. The first one depends on the optimization problem to be solved, $H_{P}$, while the second is problem-independent and depends on the so-called mixer Hamiltonian, $H_M$, which is the sum of the \texttt{NOT} gates applied to all qubits: $H_M = \sum_{j=1}^{n} X_j$. The role of $U(H_{M}, \beta_k)$ is to increase the probability of measuring high-quality solutions. The QAOA ansatz is given by the following equation 
\begin{equation}
    \label{eq:qaoa}
    \ket{\psi_{(\gamma,\beta)}} = \left( \prod_{k=p}^{1} U(H_{M}, \beta_{k}) U(H_{P}, \gamma_{k}) \right)  \, \, \, H^{\otimes n} \ket{0} \; ,
\end{equation}
where the problem Hamiltonian $H_{P}$ is a diagonal matrix where the eigenvalues are the values of the objective function: $H_P \ket{x} = f(x) \ket{x}$. The symbol $H^{\otimes n}$ represents the application of a Hadamard gate to each qubit. Observe that in Equation~\eqref{eq:qaoa} the operators are written in the opposite order compared to the circuit (see Figure~\ref{fig:QAOA}). The vectors $\beta_k$ and $\gamma_{k}$ are the parameters of the ansatz, that will be optimized by a classical optimizer to maximize the expected value of $H_P$ in state $\ket{\psi}$:
\begin{equation}
\label{eq:ev}
    E_{(\gamma,\beta)} = \bra{\psi_{(\gamma,\beta)}} H_P \ket{\psi_{(\gamma,\beta)}}.
\end{equation}

One important question at this point is how to express the objective function $f(x)$ as a diagonal operator $H_P$ to be included in the ansatz. In order to do this we will use the only non-trivial diagonal quantum gate we have seen: the {Pauli-Z} gate. We already saw in Section~\ref{sec:background} that $Z_j\ket{x_{j}} = (-1)^{x_{j}}\ket{x_{j}}$. Thus, we only need to express $f(x)$ in terms of $s_j=(-1)^{x_j}$ variables, which take values in $\{-1,1\}$. This is just a generalized Ising model. We can do the substitution $x_{j}$ = ${(1 - s_{j})}/{2}$ in $f(x)$ to obtain the generalized Ising model and then replace the $s_j$ variables by the $Z_j$ gate to obtain the problem Hamiltonian.

\begin{example}
\label{ex:hp}
Let us build the problem Hamiltonian for the pseudo-Boolean function at the end of Example~\ref{ex:integer-function}. We get this Hamiltonian by replacing $x_j$ by operator $(I-Z_j)/2$:
    \begin{align*}
        H_{P} & =-9+13 \frac{(I - Z_{1})}{2} +14 \frac{(I - Z_{2})}{2} + 9 \frac{(I - Z_{3})}{2} \\
        &-18 \frac{(I - Z_{1})}{2} \otimes \frac{(I - Z_{2})}{2} -18 \frac{(I - Z_{1})}{2} \otimes \frac{(I - Z_{3})}{2} -18 \frac{(I - Z_{2})}{2} \otimes\frac{(I - Z_{3})}{2} \\
        &+36 \frac{(I - Z_{1})}{2} \otimes\frac{(I - Z_{2})}{2}  \otimes\frac{(I - Z_{3})}{2} .
        \end{align*}

Expanding the tensor products we get
        \begin{align}
        \label{eqn:hp-example-eq}
        H_P &= -2 Z_1 - \frac{5}{2} Z_2 - \frac{9}{2} Z_1 \otimes Z_2 \otimes Z_3 .
        \end{align}
    \qed
\end{example}

The problem Hamiltonian $H_P$ will have the general form
\begin{equation}
    \label{eqn:generalized-hamiltonian}
    H_P = \sum_{S \subseteq [n]} a_S \bigotimes_{j \in S} Z_j,
\end{equation}
and the unitary transformation $U(H_{P}, \gamma_{k})$ can be written as
\begin{equation}
    U(H_{P}, \gamma_{k}) = e^{-i \gamma_k \sum_{S \subseteq [n]} a_S \bigotimes_{j \in S} Z_j}
    = \prod_{S \subseteq [n]} e^{-i \gamma_k a_S \bigotimes_{j \in S} Z_j},
\end{equation}
where we use the fact that the $Z_j$ operators and their tensor products commute to transform the exponential of a sum into the product of exponentials. In order to build the circuit associated to $U(H_{P}, \gamma_{k})$ we need to express the terms $e^{-i \gamma_k a_S \bigotimes_{j \in S} Z_j}$ with quantum gates.

Let us start with the linear terms in the problem Hamiltonian, that is terms with the form $e^{-i\gamma_{k}h_{j}Z_{j}}$. 
We saw in Section~\ref{sec:background} that the exponential of a diagonal matrix is a diagonal matrix. In the case of the linear terms we have
\begin{equation}\label{eq:hpu}
e^{-i\gamma_{k}h_{j}Z_{j}} =
\begin{pmatrix}
e^{-i\gamma_{k}h_{j}} & 0 \\
0 & e^{i\gamma_{k}h_{j}}
\end{pmatrix}    
= 
R_{Z_{j}}(2\gamma_{k}h_{j}),
\end{equation}
where $R_{Z_{j}}(2\gamma_{k}h_{j})$ is the \texttt{RZ} quantum gate with angle $2\gamma_{k}h_{j}$ applied to qubit $j$.

Let us now focus on the second order terms $e^{-i \gamma_{k} J_{j\ell} Z_{j} \otimes Z_{\ell}}$. Following the same expansion of the exponential we get
\begin{equation}\label{eq:hpu2}
\resizebox{1\hsize}{!}{$
Z_{j} \otimes Z_{\ell} =
\begin{pmatrix}
1 & 0 & 0 & 0\\
0 & -1 & 0 & 0 \\
0 & 0 & -1 & 0 \\
0 & 0 & 0 & 1
\end{pmatrix}    
\Rightarrow
e^{-i \gamma_{k} J_{j\ell} Z_{j} \otimes Z_{\ell}} =
\begin{pmatrix}
e^{-i\gamma_{k}J_{j\ell}} & 0 & 0 & 0 \\
0 & e^{i\gamma_{k}J_{j\ell}} & 0 & 0 \\
0 & 0 & e^{i\gamma_{k}J_{j\ell}} & 0 \\
0 & 0 & 0 & e^{-i\gamma_{k}J_{j\ell}} 
\end{pmatrix}
= 
R_{Z_{j}Z_{\ell}}(2\gamma_{k}J_{j\ell})$},
\end{equation}
where we use the \texttt{RZZ} quantum gate that is present in some quantum computers and simulators. 
The \texttt{RZZ} quantum gate can be expressed in terms of the \texttt{RZ} and \texttt{CNOT} gates as follows
\begin{equation} \label{eq:zz}
R_{Z_{j}Z_{\ell}}(\alpha) = \texttt{CNOT}_{j\ell} \; R_{Z_{\ell}}(\alpha) \; \texttt{CNOT}_{j\ell} .
\end{equation}

In general, an exponential term $e^{-i \gamma_{k} a_S \bigotimes_{j \in S} Z_{j}}$ can be implemented in a quantum computer with the circuit
\begin{equation}
    e^{-i \gamma_{k} a_S \bigotimes_{j \in S} Z_{j}} = 
    \prod_{j=1}^{|S|-1} \texttt{CNOT}_{S_{j},S_{j+1}} 
    R_{Z_{S_{|S|}}}(2 \gamma_k a_S)
    \prod_{j=|S|-1}^{1} \texttt{CNOT}_{S_{j},S_{j+1}},
\end{equation}
where we assumed that $S$ is a set with ordered elements and $S_j$ represents the $j$-th element in that ordered set.

\begin{example}
\label{ex:circuit-hp}
Let us implement a circuit for $U(H_P, \gamma_k)$ where $H_P$ is given by Equation~\eqref{eqn:hp-example-eq} of Example~\ref{ex:hp}. There is a cubic and two linear terms in the expression for $H_P$ and using \texttt{CNOT} and \texttt{RZ} gates we obtain the circuit in Figure~\ref{fig:circuit-hp}.
\begin{figure}
    \centering
    \begin{quantikz}
    \lstick{$q_1$} & \gate{RZ(-4 \gamma_k)} & \ctrl{1} & \qw      & \qw                    
    & \qw      & \ctrl{1} & \qw \\
    \lstick{$q_2$} & \gate{RZ(-5 \gamma_k)} & \targ{}  & \ctrl{1} & \qw                    
    & \ctrl{1} & \targ{} & \qw  \\
    \lstick{$q_3$} & \qw                    & \qw      & \targ{}  & \gate{RZ(-9 \gamma_k)} 
    & \targ{}  & \qw & \qw 
    \end{quantikz}
    \caption{Circuit for $U(H_P, \gamma_k)$ where $H_P=-2 Z_1 - \frac{5}{2} Z_2 - \frac{9}{2} Z_1 \otimes Z_2 \otimes Z_3$.}
    \label{fig:circuit-hp}
\end{figure}
\qed
\end{example}

The last component of the ansatz we need to implement is the mixer $U(H_{M}, \beta_{k})$, whose expression is 
\begin{equation} \label{eq:umh}
    U(H_{M}, \beta_{k}) = e^{-i \beta_{k} \sum_{j=1}^{n}X_{j}} = \prod_{j=1}^{n} e^{-i \beta_{k} X_{j}},
\end{equation}
where again we used the commutativity of the \text{NOT} gates to transform the exponential of a sum into a product of exponentials. The building blocks of the previous expression are the terms $e^{-i \beta_{k} X_{j}}$, which are developed as follows
\begin{equation}\label{eq:gumh}
e^{-i\beta_{k}X_{j}} =
\begin{pmatrix}
\cos \beta_{k} & -i \sin \beta_{k} \\
-i  \sin \beta_{k} & \cos \beta_{k}
\end{pmatrix}    
= 
R_{X_{j}}(2\beta_{k}),
\end{equation}
where $R_{X_{j}}(2\beta_{k})$ is the \texttt{RX} gate with angle $2\beta_{k}$ for qubit $j$.

\begin{example}
    \label{ex:circuit-hm}
    In a three-qubit asatz, like the one required for Example~\ref{ex:hp}, the mixer $U(H_{M}, \beta_{k})$ is given by the circuit in Figure~\ref{fig:circuit-hm}. \qed
    \begin{figure}
    \centering
    \begin{quantikz}
    \lstick{$q_1$} & \gate{RX(2\beta_k)} & \qw \\
    \lstick{$q_2$} & \gate{RX(2\beta_k)} & \qw \\
    \lstick{$q_3$} & \gate{RX(2\beta_k)} & \qw \\
    \end{quantikz}
    \caption{Circuit for $U(H_M, \beta_k)$ for the ansatz required to optimize the problem in Examples~\ref{ex:hp} and~\ref{ex:circuit-hp}.}
    \label{fig:circuit-hm}
\end{figure}
\qed
\end{example}

Qiskit is the IBM SDK to simulate and run quantum circuits in IBM gate-based quantum computers. We show in Figure~\ref{fig:qaoa-main} the main code to minimize the integer function in Example~\ref{ex:integer-function}. The COBYLA algorithm has been used to minimize the objective function in the classical computer (Line 17). The code to build the quantum circuit is in Figure~\ref{fig:qaoa-ansatz}. It follows the design obtained in Examples~\ref{ex:circuit-hp} and~\ref{ex:circuit-hm}. Finally, the code to run the circuit, sample the solutions and compute the average objective value for the sampling is shown in Figure~\ref{fig:qaoa-sampling}. Observe that this code needs to evaluate the function using the original expression in Example~\ref{ex:integer-function}, and not the problem Hamiltonian of Example~\ref{ex:circuit-hp}. In Table~\ref{tab:qaoa-results} we show the results of a run of the code. We observe that the optimal solution ($000$) is the most frequent solution we obtain, but thee other suboptimal solutions with a high number of occurrences.

\begin{figure}[!ht]
    \begin{Verbatim}[commandchars=\\\{\},numbers=left,firstnumber=1,stepnumber=1,fontsize=\scriptsize,frame=lines]
\PYG{c+c1}{\PYGZsh{} Import Necessary Packages: General and Quantum Simulation}
\PYG{k+kn}{from} \PYG{n+nn}{qiskit.circuit} \PYG{k+kn}{import} \PYG{n}{QuantumCircuit}
\PYG{k+kn}{from} \PYG{n+nn}{qiskit.visualization} \PYG{k+kn}{import} \PYG{n}{plot\PYGZus{}histogram}
\PYG{k+kn}{from} \PYG{n+nn}{qiskit\PYGZus{}aer} \PYG{k+kn}{import} \PYG{n}{Aer}
\PYG{k+kn}{from} \PYG{n+nn}{scipy.optimize} \PYG{k+kn}{import} \PYG{n}{minimize}

\PYG{c+c1}{\PYGZsh{} The QAOA\PYGZsq{}s Ansatz Wrapper}
\PYG{k}{def} \PYG{n+nf}{build\PYGZus{}ansatz}\PYG{p}{(}\PYG{n}{theta}\PYG{p}{):}
    \PYG{c+c1}{\PYGZsh{} Omitted}

\PYG{c+c1}{\PYGZsh{} Classical objective function}
\PYG{k}{def} \PYG{n+nf}{classical\PYGZus{}obj\PYGZus{}fn}\PYG{p}{(}\PYG{n}{theta}\PYG{p}{,} \PYG{n}{num\PYGZus{}executions}\PYG{o}{=}\PYG{l+m+mi}{1000}\PYG{p}{):}
    \PYG{c+c1}{\PYGZsh{} Omitted}

\PYG{c+c1}{\PYGZsh{} The main program: Simulation, Parameters\PYGZsq{} and Solutions\PYGZsq{} Optimisation}
\PYG{n}{quantum\PYGZus{}device} \PYG{o}{=} \PYG{n}{Aer}\PYG{o}{.}\PYG{n}{get\PYGZus{}backend}\PYG{p}{(}\PYG{l+s+s1}{\PYGZsq{}qasm\PYGZus{}simulator\PYGZsq{}}\PYG{p}{)}
\PYG{n}{qaoa\PYGZus{}optimized\PYGZus{}parameters} \PYG{o}{=} \PYG{n}{minimize}\PYG{p}{(}\PYG{n}{classical\PYGZus{}obj\PYGZus{}fn}\PYG{p}{,} \PYG{p}{[}\PYG{l+m+mf}{0.1}\PYG{p}{,}\PYG{l+m+mf}{0.1}\PYG{p}{,}\PYG{l+m+mf}{0.1}\PYG{p}{,}\PYG{l+m+mf}{0.1}\PYG{p}{],} \PYG{n}{method}\PYG{o}{=}\PYG{l+s+s1}{\PYGZsq{}COBYLA\PYGZsq{}}\PYG{p}{)} 
\PYG{k}{print}\PYG{p}{(}\PYG{n}{f}\PYG{l+s+s1}{\PYGZsq{}Parameters: \PYGZob{}qaoa\PYGZus{}optimized\PYGZus{}parameters\PYGZcb{}\PYGZsq{}}\PYG{p}{)}
\PYG{n}{final\PYGZus{}circuit} \PYG{o}{=} \PYG{n}{build\PYGZus{}ansatz}\PYG{p}{(}\PYG{n}{qaoa\PYGZus{}optimized\PYGZus{}parameters}\PYG{o}{.}\PYG{n}{x}\PYG{p}{)}
\PYG{n}{sampled\PYGZus{}solutions} \PYG{o}{=} \PYG{n}{quantum\PYGZus{}device}\PYG{o}{.}\PYG{n}{run}\PYG{p}{(}\PYG{n}{final\PYGZus{}circuit}\PYG{p}{,} \PYG{n}{nshots} \PYG{o}{=} \PYG{l+m+mi}{1000}\PYG{p}{)}\PYG{o}{.}\PYGZbs{}
                        \PYG{n}{result}\PYG{p}{()}\PYG{o}{.}\PYG{n}{get\PYGZus{}counts}\PYG{p}{()}
\PYG{n}{plot\PYGZus{}histogram}\PYG{p}{(}\PYG{n}{sampled\PYGZus{}solutions}\PYG{p}{)}
\end{Verbatim}
    \caption{Main code of the QAOA implementation in Qiskit of the program to minimize the objective function of Example~\ref{ex:integer-function}.}
    \label{fig:qaoa-main}
\end{figure}

\begin{figure}[!hp]
    \begin{Verbatim}[commandchars=\\\{\},numbers=left,firstnumber=1,stepnumber=1,fontsize=\scriptsize,frame=lines]
\PYG{c+c1}{\PYGZsh{} The QAOA\PYGZsq{}s Ansatz Wrapper}
\PYG{k}{def} \PYG{n+nf}{build\PYGZus{}ansatz}\PYG{p}{(}\PYG{n}{theta}\PYG{p}{):}
  \PYG{c+c1}{\PYGZsh{} Parameters of the QAOA\PYGZsq{}s Ansatz}
  \PYG{n}{num\PYGZus{}qubits} \PYG{o}{=} \PYG{l+m+mi}{3}
  \PYG{n}{qaoa\PYGZus{}depth} \PYG{o}{=} \PYG{n+nb}{len}\PYG{p}{(}\PYG{n}{theta}\PYG{p}{)}\PYG{o}{//}\PYG{l+m+mi}{2}
  \PYG{c+c1}{\PYGZsh{} Construct the QAOA\PYGZsq{}s Circuit}
  \PYG{n}{circuit} \PYG{o}{=} \PYG{n}{QuantumCircuit}\PYG{p}{(}\PYG{n}{num\PYGZus{}qubits}\PYG{p}{)}
  \PYG{n}{gamma} \PYG{o}{=} \PYG{n}{theta}\PYG{p}{[:}\PYG{n}{qaoa\PYGZus{}depth}\PYG{p}{]}
  \PYG{n}{beta} \PYG{o}{=} \PYG{n}{theta}\PYG{p}{[}\PYG{n}{qaoa\PYGZus{}depth}\PYG{p}{:]}
  \PYG{c+c1}{\PYGZsh{} Layer of Hadamard gates}
  \PYG{k}{for} \PYG{n}{i} \PYG{o+ow}{in} \PYG{n+nb}{range}\PYG{p}{(}\PYG{l+m+mi}{0}\PYG{p}{,} \PYG{n}{num\PYGZus{}qubits}\PYG{p}{):}
    \PYG{n}{circuit}\PYG{o}{.}\PYG{n}{h}\PYG{p}{(}\PYG{n}{i}\PYG{p}{)}
  \PYG{c+c1}{\PYGZsh{} Construct QAOA layers}
  \PYG{k}{for} \PYG{n}{i} \PYG{o+ow}{in} \PYG{n+nb}{range}\PYG{p}{(}\PYG{l+m+mi}{0}\PYG{p}{,}\PYG{n}{qaoa\PYGZus{}depth}\PYG{p}{):}
    \PYG{c+c1}{\PYGZsh{} Unitary depending on problem Hamiltonian}
    \PYG{n}{circuit}\PYG{o}{.}\PYG{n}{rz}\PYG{p}{(}\PYG{o}{\PYGZhy{}}\PYG{l+m+mi}{4} \PYG{o}{*} \PYG{n}{gamma}\PYG{p}{[}\PYG{n}{i}\PYG{p}{],}\PYG{l+m+mi}{0}\PYG{p}{)}
    \PYG{n}{circuit}\PYG{o}{.}\PYG{n}{rz}\PYG{p}{(}\PYG{o}{\PYGZhy{}}\PYG{l+m+mi}{5} \PYG{o}{*} \PYG{n}{gamma}\PYG{p}{[}\PYG{n}{i}\PYG{p}{],}\PYG{l+m+mi}{1}\PYG{p}{)}
    \PYG{n}{circuit}\PYG{o}{.}\PYG{n}{cx}\PYG{p}{(}\PYG{l+m+mi}{0}\PYG{p}{,}\PYG{l+m+mi}{1}\PYG{p}{)}
    \PYG{n}{circuit}\PYG{o}{.}\PYG{n}{cx}\PYG{p}{(}\PYG{l+m+mi}{1}\PYG{p}{,}\PYG{l+m+mi}{2}\PYG{p}{)}
    \PYG{n}{circuit}\PYG{o}{.}\PYG{n}{rz}\PYG{p}{(}\PYG{o}{\PYGZhy{}}\PYG{l+m+mi}{9} \PYG{o}{*} \PYG{n}{gamma}\PYG{p}{[}\PYG{n}{i}\PYG{p}{],}\PYG{l+m+mi}{2}\PYG{p}{)}
    \PYG{n}{circuit}\PYG{o}{.}\PYG{n}{cx}\PYG{p}{(}\PYG{l+m+mi}{1}\PYG{p}{,}\PYG{l+m+mi}{2}\PYG{p}{)}
    \PYG{n}{circuit}\PYG{o}{.}\PYG{n}{cx}\PYG{p}{(}\PYG{l+m+mi}{0}\PYG{p}{,}\PYG{l+m+mi}{1}\PYG{p}{)}
    \PYG{c+c1}{\PYGZsh{} Construct the mixer}
    \PYG{k}{for} \PYG{n}{qbit} \PYG{o+ow}{in} \PYG{n+nb}{range}\PYG{p}{(}\PYG{l+m+mi}{0}\PYG{p}{,}\PYG{n}{num\PYGZus{}qubits}\PYG{p}{):}
      \PYG{n}{circuit}\PYG{o}{.}\PYG{n}{rx}\PYG{p}{(}\PYG{l+m+mi}{2} \PYG{o}{*} \PYG{n}{beta}\PYG{p}{[}\PYG{n}{i}\PYG{p}{],} \PYG{n}{qbit}\PYG{p}{)}
  \PYG{c+c1}{\PYGZsh{} Apply measurement to all qubits}
  \PYG{n}{circuit}\PYG{o}{.}\PYG{n}{measure\PYGZus{}all}\PYG{p}{()}
  \PYG{k}{return} \PYG{n}{circuit}
\end{Verbatim}
    \caption{Code to build the ansatz of QAOA for Example~\ref{ex:integer-function}.}
    \label{fig:qaoa-ansatz}
\end{figure}

\begin{figure}[!hp]
    \begin{Verbatim}[commandchars=\\\{\},numbers=left,firstnumber=1,stepnumber=1,fontsize=\scriptsize,frame=lines]
\PYG{c+c1}{\PYGZsh{} Classical objective function}
\PYG{k}{def} \PYG{n+nf}{classical\PYGZus{}obj\PYGZus{}fn}\PYG{p}{(}\PYG{n}{theta}\PYG{p}{,} \PYG{n}{num\PYGZus{}executions}\PYG{o}{=}\PYG{l+m+mi}{1000}\PYG{p}{):}
  \PYG{c+c1}{\PYGZsh{} Objective function to evaluate after sampling}
  \PYG{k}{def} \PYG{n+nf}{objective\PYGZus{}function}\PYG{p}{(}\PYG{n}{x}\PYG{p}{):}
    \PYG{k}{return} \PYG{o}{\PYGZhy{}}\PYG{l+m+mi}{9} \PYG{o}{+}\PYG{l+m+mi}{13}\PYG{o}{*}\PYG{n}{x}\PYG{p}{[}\PYG{l+m+mi}{0}\PYG{p}{]} \PYG{o}{+}\PYG{l+m+mi}{14}\PYG{o}{*}\PYG{n}{x}\PYG{p}{[}\PYG{l+m+mi}{1}\PYG{p}{]} \PYG{o}{+}\PYG{l+m+mi}{9}\PYG{o}{*}\PYG{n}{x}\PYG{p}{[}\PYG{l+m+mi}{2}\PYG{p}{]} \PYGZbs{}
           \PYG{o}{\PYGZhy{}}\PYG{l+m+mi}{18}\PYG{o}{*}\PYG{n}{x}\PYG{p}{[}\PYG{l+m+mi}{0}\PYG{p}{]}\PYG{o}{*}\PYG{n}{x}\PYG{p}{[}\PYG{l+m+mi}{1}\PYG{p}{]} \PYG{o}{\PYGZhy{}}\PYG{l+m+mi}{18}\PYG{o}{*}\PYG{n}{x}\PYG{p}{[}\PYG{l+m+mi}{0}\PYG{p}{]}\PYG{o}{*}\PYG{n}{x}\PYG{p}{[}\PYG{l+m+mi}{2}\PYG{p}{]} \PYG{o}{\PYGZhy{}}\PYG{l+m+mi}{18}\PYG{o}{*}\PYG{n}{x}\PYG{p}{[}\PYG{l+m+mi}{1}\PYG{p}{]}\PYG{o}{*}\PYG{n}{x}\PYG{p}{[}\PYG{l+m+mi}{2}\PYG{p}{]} \PYGZbs{}
           \PYG{o}{+}\PYG{l+m+mi}{36}\PYG{o}{*}\PYG{n}{x}\PYG{p}{[}\PYG{l+m+mi}{0}\PYG{p}{]}\PYG{o}{*}\PYG{n}{x}\PYG{p}{[}\PYG{l+m+mi}{1}\PYG{p}{]}\PYG{o}{*}\PYG{n}{x}\PYG{p}{[}\PYG{l+m+mi}{2}\PYG{p}{]}

  \PYG{c+c1}{\PYGZsh{} Create the Ansatz}
  \PYG{n}{circuit} \PYG{o}{=} \PYG{n}{build\PYGZus{}ansatz}\PYG{p}{(}\PYG{n}{theta}\PYG{p}{)}
  \PYG{c+c1}{\PYGZsh{} Execute the Circuit}
  \PYG{n}{sampled\PYGZus{}solutions} \PYG{o}{=} \PYG{n}{quantum\PYGZus{}device}\PYG{o}{.}\PYG{n}{run}\PYG{p}{(}\PYG{n}{circuit}\PYG{p}{,} \PYG{n}{nshots} \PYG{o}{=} \PYG{n}{num\PYGZus{}executions}\PYG{p}{)}\PYG{o}{.}\PYGZbs{}
                        \PYG{n}{result}\PYG{p}{()}\PYG{o}{.}\PYG{n}{get\PYGZus{}counts}\PYG{p}{()}
  \PYG{c+c1}{\PYGZsh{} Compute the average value of the objective function for sampled solutions}
  \PYG{n}{value} \PYG{o}{=} \PYG{l+m+mi}{0}
  \PYG{n}{sampled} \PYG{o}{=} \PYG{l+m+mi}{0}
  \PYG{k}{for} \PYG{n}{solution}\PYG{p}{,} \PYG{n}{occurrences} \PYG{o+ow}{in} \PYG{n}{sampled\PYGZus{}solutions}\PYG{o}{.}\PYG{n}{items}\PYG{p}{():}
    \PYG{c+c1}{\PYGZsh{} Invert and transform the solution because it is a string}
    \PYG{c+c1}{\PYGZsh{} Element 0 is at position n\PYGZhy{}1, element 1 is at posicion n\PYGZhy{}2, etc.}
    \PYG{n}{x}\PYG{o}{=}\PYG{p}{[}\PYG{n+nb}{int}\PYG{p}{(}\PYG{n}{v}\PYG{p}{)} \PYG{k}{for} \PYG{n}{v} \PYG{o+ow}{in} \PYG{n}{solution}\PYG{p}{[::}\PYG{o}{\PYGZhy{}}\PYG{l+m+mi}{1}\PYG{p}{]]}
    \PYG{n}{value} \PYG{o}{=} \PYG{n}{objective\PYGZus{}function}\PYG{p}{(}\PYG{n}{x}\PYG{p}{)} \PYG{o}{*} \PYG{n}{occurrences}
    \PYG{n}{sampled} \PYG{o}{+=} \PYG{n}{occurrences}
  \PYG{k}{return} \PYG{n}{value}\PYG{o}{/}\PYG{n}{sampled} 
\end{Verbatim}
    \caption{Code to sample the ansatz and compute the average objective function over the sampling for Example~\ref{ex:integer-function}.}
    \label{fig:qaoa-sampling}
\end{figure}

\begin{table}[!ht]
\centering
\caption{Results obtained after running QAOA using the code in Figures~\ref{fig:qaoa-main} to~\ref{fig:qaoa-sampling}.}
\label{tab:qaoa-results}
    \begin{tabular}{lrclr}
    \toprule
    Solution & Occurrences & \phantom{aa} & Solution & Occurrences\\
    \cline{1-2} \cline{4-5}
    000 & 248 & & 100 &  36 \\
    001 &  52 & & 101 & 199 \\
    010 &   9 & & 110 & 169 \\
    011 & 200 & & 111 & 111 \\
    \bottomrule
    \end{tabular}
\end{table}

\section{Discussion and Future Directions}
\label{sec:conclusions}

We have shown in this paper how to solve combinatorial optimization problems using quantum computing. We have provided a general and systematic approach for the transformation of the original problem into an equivalent problem that is ready to be optimized by the current quantum computers. When solving one particular problem, it is possible to focus on some properties of the problem to simplify steps, reduce the number of added variables or estimate the penalty constants~\cite{DBLP:conf/gecco/GohBGL22}. More research needs to be done to understand how to fix the penalty constants and how to minimize the number of variables that we need to add to quadratize the objective function~\cite{DBLP:journals/corr/abs-1910-13583}.

There are many different ways to transform a combinatorial optimization problem, and different transformations can yield different performance. 
The recent literature contains examples of this impact~\cite{DBLP:conf/gecco/ZielinskiNSGLF23}. It is an open question which transformation is the most appropriate for a given problem.

We assumed along this paper that we have a mathematical expression of the objective function. This is not always the case. The objective function could be a simulation in a classical computer, which makes it difficult (or impossible) to transform the problem into a mathematical expression. There are proposals like AutoQUBO~\cite{DBLP:conf/gecco/MoraglioGS22,DBLP:conf/gecco/PauckertAGGP23} to find the QUBO associated to an arbitrary black-box objective function.

While we focus here on single-objective optimization, we can also find a few examples in the literature where multi-objective optimization is the target, either using quantum annealers~\cite{DBLP:conf/gecco/Ayodele00LP23} or gate-based computers~\cite{DBLP:conf/ppsn/DahiCLDA24}. At the time of writing there are less than 5 papers on this topic in the current literature, so we can say that multi-objective optimization using quantum computers is an emerging research area.

Finally, if we focus on more fundamental aspects of quantum optimization, it is not yet clear if quantum computers can optimize functions faster in practice than classical computers. In fact, there are some works that prove that it is possible to efficiently evaluate the average value of the objective function for solutions generated using the ansatz of QAOA in a classical machine~\cite{DBLP:conf/gecco/ChicanoDL23}. Of course, this is not solving the problem, but it suggests that the parameter tuning of the ansatz in QAOA can be done in a classical machine if the number of layers $p$ is low enough; using the quantum computer only in the very last sampling of QAOA to obtain the solutions. The same work also suggests a quantum-inspired algorithm based on this result. This raises the question if QAOA can optimize problems faster than classical algorithms, which is still unanswered. Furthermore, in a recent paper~\cite{HoeflerHT23} the authors conclude that current gate-based quantum computers cannot compete with GPUs and the quadratic speedup provided by Grover algorithm and amplitude amplification are not enough to solve search and optimization problems faster than the current classical hardware. This means we need new algorithmic approaches. The quadratic speedup is the maximum speedup we can obtain using a black-box approach to optimization~\cite{BennettBBV97}. This means that we need to use \emph{gray-box}~\cite{DBLP:conf/ppsn/ChicanoWOT24} and \emph{white-box} approaches in order to get quantum supremacy in optimization.

\section*{Acknowledgements}
This research is partially funded by project PID 2020-116727RB-I00 (HUmove) funded by MCIN/AEI/ 10.13039/501100011033; TAILOR ICT-48 Network (No 952215) funded by EU Horizon 2020 research and innovation programme; Junta de Andalucia, Spain, under contract QUAL21 010UMA; and the University of Malaga (PAR 4/2023).


\end{document}